\documentclass{article}
\begin{document}

\title{A Tractable Complex Network Model Based on the Stochastic Mean-Field
Model of Distance}
\author{David J. Aldous\\
Department of Statistics\\
367 Evans Hall\\
U.C. Berkeley CA 94720. \\
\texttt{aldous@stat.berkeley.edu}}
\newtheorem{Lemma}{Lemma}
\newtheorem{Proposition}[Lemma]{Proposition}
\newtheorem{Algorithm}[Lemma]{Algorithm}
\newtheorem{Theorem}[Lemma]{Theorem}
\newtheorem{Conjecture}[Lemma]{Conjecture}
\newtheorem{Problem}[Lemma]{Problem}
\newtheorem{Example}[Lemma]{Example}
\newtheorem{OP}[Lemma]{Open Problem}
\newtheorem{Hypothesis}[Lemma]{Hypothesis}
\newtheorem{Construction}[Lemma]{Construction}
\newtheorem{Corollary}[Lemma]{Corollary}

\newcommand{\con}{\rightarrow}
\newcommand{\ed}{\ \stackrel{d}{=} \ }
\newcommand{\cd}{\ \stackrel{d}{\rightarrow} \ }
\newcommand{\cp}{\ \stackrel{p}{\rightarrow} \ }
\newcommand{\TT}{\mbox{${\cal T}$}}
\newcommand{\EE}{\mbox{${\cal E}$}}
\newcommand{\VV}{\mbox{${\cal V}$}}
\newcommand{\GG}{\mbox{${\cal G}$}}
\newcommand{\LL}{\mbox{${\cal L}$}}
\newcommand{\FF}{\mbox{${\cal F}$}}
\renewcommand{\SS}{\mbox{${\cal S}$}}
\newcommand{\NN}{\mbox{${\cal N}$}}
\newcommand{\eps}{\varepsilon}
\newcommand{\bA}{{\bf A}}
\newcommand{\bB}{{\bf B}}
\newcommand{\bE}{{\bf E}}
\newcommand{\bP}{{\bf P}}
\newcommand{\bR}{{\bf R}}
\newcommand{\bZ}{{\bf Z}}
\newcommand{\bJ}{{\bf J}}
\newcommand{\bQ}{{\bf Q}}
\newcommand{\bX}{{\bf X}}
\newcommand{\bY}{{\bf Y}}
\newcommand{\bV}{{\bf V}}
\newcommand{\bx}{{\bf x}}
\newcommand{\bp}{{\bf p}}
\newcommand{\by}{{\bf y}}
\newcommand{\ED}{\stackrel{\con}{{\cal E}} }
\newcommand{\var}{{\rm var}\ }
\newcommand{\dens}{{\rm dens}}
\newcommand{\bdens}{\stackrel{\_\_\_\_}{\dens}}
\newcommand{\Ag}{||g||_2^2}
\newcommand{\sfrac}[2]{{\textstyle\frac{#1}{#2}}}

\newcommand{\Din}{D_{\mbox{{\small in}}}}
\newcommand{\Dout}{D_{\mbox{{\small out}}}}
\newcommand{\cbluster}{\kappa_{\mbox{{\small cluster}}}}
\newcommand{\ccluster}{\kappa_{\mathrm{cluster}}}
\newcommand{\Geo}{\mbox{Geo}}
\newcommand{\Exp}{\mbox{Exp}}
\newcommand{\Bin}{\mbox{Bin}}
\newcommand{\Poi}{\mbox{Poi}}
\newcommand{\Ber}{\mbox{Ber}}
\newcommand{\bD}{{\bf D}}
\newcommand{\bxi}{\Xi}
\newcommand{\dmetric}{d_{\mbox{\tiny{metric}}}}
\newcommand{\dgraph}{d_{\mbox{\tiny{graph}}}}
\newcommand{\low}{\mbox{[low]}}
\newcommand{\high}{\mbox{[high]}}
\newcommand{\nn}{near neighbor}
\renewcommand{\root}{\mbox{root}}
\newcommand{\DD}{\mathcal{D}}
\newcommand{\CC}{\mathcal{C}}
\newcommand{\bchi}{\bar{\chi}}
\maketitle

\begin{abstract}
Much recent research activity has been devoted to empirical study
and theoretical models of complex networks (random graphs) possessing three
qualitative features: power-law degree distributions, local clustering, and
slowly-growing diameter.
We point out a new (in this context) platform for such models -- the
stochastic mean-field model of distances -- and within this platform
study a simple two-parameter proportional attachment (or copying) model.
The model is mathematically natural, permits a wide variety
of explicit calculations, has the desired three qualitative features,
and fits the complete range of degree scaling exponents and clustering parameters;
in these respects it compares favorably with existing models.
\end{abstract}

\section{Introduction}
\label{sec-INT}
The topic of {\em complex networks}, more precisely the design and
theoretical analysis of stochastic models of large graphs which differ
from the classical
Erd{\H{o}}s - R{\'e}nyi model, has attracted intense recent attention,
surveyed from a statistical physics viewpoint in
\cite{AB02,DM02,newman-survey} and from a rigorous mathematical viewpoint in
\cite{boll-scale}.

Let us frame one aspect of this topic, by analogy.
In freshman statistics we learn that bivariate data
(e.g. heights and weights of $n$ individuals)
can be summarized by $5$ {\em summary statistics}:
average height, standard deviation of height, average weight, standard deviation of weight,
correlation coefficient.
And there is a $5$-parameter probability model, the bivariate Normal,
which (in several precise senses) exactly corresponds to these
particular summary statistics.
In the context of real-world graphs
(where we will always regard the number $n$ of vertices as large),
one could analogously seek a crude statistical description by reporting a set of
summary statistics.
An evident choice is
\begin{itemize}
\item $\partial = $ average vertex-degree
\end{itemize}
and recently popular extra choices include
\begin{itemize}
\item an exponent $\gamma$ characterizing power-law tail behavior of degree distribution
\item a ``clustering coefficient" $\kappa$ measuring relative density of triangles
\item the average distance $\bar{\ell}$ between vertex-pairs.
\end{itemize}
These choices reflect and seek to quantify three qualitative features claimed to hold in many
interesting graphs (from WWW links to human social networks): 
power-law degree distribution, local clustering of edges, and diameter growing
as $O(\log n)$. 
So from the viewpoint of classical mathematical statistics,
it would be natural to seek a several-parameter stochastic model
of random graphs whose parameters could be readily identified
with summary statistics of the kind above.
In more detail, we propose three desiderata\footnote{From an applied viewpoint,
one could regard ``fitting empirical data" as the single criterion; 
we are of course taking a theoretical viewpoint} for 
a satisfactory model,
beyond possessing the three qualitative features mentioned above:
\begin{itemize}
\item 
{\em mathematical tractability}: one can find reasonably explicit
formulas for a variety of quantities of interest
\item 
{\em fitting flexibility}: 
by varying model parameters one can vary 
summary statistics 
(like the 4 listed above)
broadly through their possible ranges
\item 
{\em naturalness}:
the qualitative properties emerge from some
simple underlying mathematical structure rather than being
forced by fiat.
\end{itemize}
Unfortunately no satisfactory such models are known.
The statistical physics literature surveyed in
\cite{AB02,DM02} 
starts with a few elementary model-construction ideas
(such as the 
{\em proportional attachment}
and 
{\em small worlds}
models mentioned in section \ref{sec-compare})
and then explores numerous variations.
Our purpose in this paper is to introduce
a new class of model 
we call
{\em metric copying},\footnote{We use
mathematical terminology: a {\em metric} is a distance function.
Confusingly, 
some engineers use ``metric" to mean ``summary statistic"}
and to study a particular two-parameter model
({\em mean-field simple copying}, MFSC)
within this class.
The description and analysis of the MFSC model
involve somewhat more sophisticated mathematical visualization
than has been used in previous complex networks literature.
So let us first address the first two desiderata
by listing results for the model (section \ref{sec-formulas}), and only later
(section \ref{sec-MODEL}) describe the model.
Section \ref{sec-CALC} derives most of the formulas in section \ref{sec-formulas},
and section \ref{sec-furtherc} exhibits further calculations.
A briefer account of the model, aimed at mathematicians, appears
in \cite{me106}.

\subsection{Some notation}
$P(\cdot)$ denotes probability,
$E(\cdot)$ denotes expectation,
and $\var(\cdot)$ denotes variance.
We assume familiarity with elementary probability
notions of random variables and their distributions.
We write
$\Geo(p), \Bin(m,p), \Exp(\mu), \Poi(\eta)$
for the geometric, binomial, exponential and Poisson distributions
in their usual parametrizations,
reviewed below.
We employ a ``blackboard shorthand" of also writing 
$\Geo(p)$ etc for a random variable with that distribution.
Thus the elementary reproductive property of the binomial
distribution could be written as
\[ \Bin(m_1+m_2,p) \ed \Bin(m_1,p) + \Bin(m_2,p) \]
where the random variables on the right are independent,
and where $\ed$ means equality in distribution.
The point of this notation is that, analogous to
``composition of functions" in which we interpret
$\exp( (x-1)^2)$ as the composition of the two functions
$\exp(x)$ and $(x-1)^2$, we can ``compose"
(statisticians say ``mix")
distributions.
For instance (cf. (\ref{Dineq}) below),
given a random variable $\Lambda$ with values in $(0,1)$
we can write $\Geo(\Lambda)$ for a random variable
whose conditional distribution given $\Lambda = p$ is the
$\Geo(p)$ distribution.

{\em Review of elementary distributions.}
\begin{eqnarray}
P(\Geo(p) = i) &=&
(1-p)^{i-1}p, \ i = 1,2,\ldots 
\nonumber \\
E \Geo(p) &=&
p^{-1}
\nonumber \\
P(\Bin(m,p) = i) &=&
{m \choose i}
p^i (1-p)^{m-i}, \ i = 0,1,\ldots,m
\nonumber \\
E \Bin(m,p) &=& mp
\label{bin-mean}
\\
\var \Bin(m,p) &=& mp(1-p)
\label{var-Bin}\\
P(\Poi(\eta) = i) &=&
e^{-\eta} \eta^i/i!
, \ i = 0,1,2,\ldots
\nonumber \\
E\Poi(\eta) &=& \eta
.\nonumber
\end{eqnarray}
The $\Exp(\mu)$ distribution has probability density function and
expectation
\begin{eqnarray}
f(x) &=& \mu e^{-\mu x}, \ 0 <x<\infty
\nonumber \\
E \Exp(\mu) &=& \mu^{-1} .
\nonumber
\end{eqnarray}
A Poisson process of rate $1$, say
$(0<\xi_1<\xi_2<\xi_3< \ldots)$,
is defined by the property
\[
\xi_1, \xi_2-\xi_1, \xi_3-\xi_2,\ldots 
\mbox{ are independent with $\Exp(1)$ distribution} \]
and has the property
\begin{equation}
P(\mbox{some } \xi_i \in [x,x+dx]) = 1 \cdot dx , \ 0 <x<\infty . \label{pp-rate}
\end{equation}

\subsection{Organization of paper}
Because the precise definition and {\em a priori} motivation of the model are
lengthy to explain,
we start by 
emphasising the {\em a posteriori} motivation, the fact that
the model permits many explicit calculations.
In first reading the formulas in section \ref{sec-formulas}, focus on the
left sides of equations, indicating what quantities can be calculated.
The formulas on the right sides will be derived in section \ref{sec-CALC}.

\section{Formulas}
\label{sec-formulas}
\subsection{Key methodology}
Like other models involving vertices arriving
and creating edges to existing vertices, the MFSC model defines a directed 
acyclic (no {\em directed} cycles) random graph $\GG_n$ on $n$ vertices.
A key feature of the model is that
there exists a well-defined limit infinite rooted graph
$\GG^*_\infty$
which represents the $n \to \infty$ limit of $\GG_n$
rooted at a uniform random (we say ``typical") vertex.
So for ``local" statistics of $\GG_n$,
one can give 
``exact formulas in the $n \to \infty$ limit" 
by doing calculations within the limit structure $\GG^*_\infty$,
and this methodology is how we will derive (section \ref{sec-CALC}) and interpret the formulas in sections \ref{sec-2par}
-- \ref{sec-EL} below.

Note that ``rooting" is introduced merely as a convenient technical way
to deal with infinite graphs.
By analogy, one could study two-dimensional space without introducing
the origin point (cf. Euclidean geometry) but for many purposes
an origin and induced coordinate system are helpful.

The MFSC model has two parameters: $\alpha, \lambda$.
In explicit formulas, 
we distinguish between a
{\em low clustering region} defined by parameter ranges
\begin{equation}
 0< \alpha < 1, \quad 0 < \lambda  \leq 1/\alpha \quad \quad \low \label{range-low} 
\end{equation}
and the complementary
{\em high clustering region} defined by $\alpha \lambda >1$;
in the latter case it is convenient to reparametrize by using
$\eta := \lambda^{-1} \log (\alpha \lambda)$
in place of $\alpha$, and the parameter ranges are
\begin{equation}
 0 < \eta < 1, \quad \eta + 1/\lambda < 1 . \quad \quad \high \label{range-high} 
\end{equation}
This distinction is purely notational; there is no
intrinsic
``non-analyticity" in the model's properties.

\subsection{The two parameters control mean degree and clustering}
\label{sec-2par}
{\bf (a).} 
First consider
$\Din$ and $\Dout$, the random in-degree and out-degree
of a typical vertex.  Then
\begin{equation}
E\Din = E \Dout (= \partial, \mbox{ say}) = \left\{ \begin{array}{cc}  \frac{\alpha}{1 - \alpha} &
\low 
\\
\frac{\eta + 1/\lambda}{1 - \eta - 1/\lambda} & \high   . 
\end{array} \right.
\label{EDeq} 
\end{equation}

\noindent
{\bf (b).}  
Second, define a normalized
{\em clustering coefficient}
$\ccluster$
in words as
\begin{quote}
The proportion of directed $2$-paths
$v_1 \to v_2 \to v_3$ 
for which $v_1 \to v_3$ is also an edge.
\end{quote}
(see (\ref{tri-form}) for a more precise definition
and derivation of (\ref{clust-form})).
Then
\begin{equation}
\ccluster = \left\{ \begin{array}{cc}  \frac{\alpha (1-\alpha) \lambda}{2 - \alpha^2 \lambda} &  \low 
\\
 \frac{(\eta + \frac{1}{2\lambda})(1 - \eta - \frac{1}{\lambda})}
{(\eta + \frac{1}{\lambda})(1- \eta - \frac{1}{2 \lambda})} & \high .
\end{array} \right. 
\label{clust-form}
\end{equation}
By solving (\ref{EDeq},\ref{clust-form})
we find (section \ref{sec-reparameter})
that every pair of values of $\partial, \ccluster$ in the complete range
\[ 0 < \partial < \infty, \quad 0 < \ccluster < 1 \]
occurs for a unique parameter pair 
$(\alpha, \lambda)$ or $(\eta, \lambda)$.
Moreover the two regions can be specified as
\begin{eqnarray*}
0 < \partial < \infty, & \ \  0 < \ccluster \leq \sfrac{1}{\partial+2} & \low \\
0 < \partial < \infty, & \ \  \sfrac{1}{\partial+2} < \ccluster < 1 & \high
\end{eqnarray*}
explaining our {\em low} and {\em high clustering} terminology.
So the two model parameters $\alpha, \lambda$ have fairly
direct interpretations in terms of mean degree and
clustering; of course we could re-parametrize the model in terms of 
$\partial$ and $\ccluster$,
but the internal mathematical structure is more conveniently
expressed using the given parameters.

\subsection{Distributions of in- and out-degrees}
{\bf (a).} 
The distribution of $\Din$ is specified as
\begin{equation}
1 + \Din \ed \Geo(e^{- \beta T})
\mbox{ where } 
T \ed \Exp(1) 
\label{Dineq}
\end{equation}
and where
\begin{equation}
\beta = \left\{ \begin{array}{cc}
\alpha & \low \\
\eta + 1/\lambda & \high .
\end{array} \right.
\label{beta-def}
\end{equation}
This works out explicitly as
\begin{equation}
P(\Din \geq d) = \frac{
\Gamma(d+1)\Gamma(1/\beta)}{\beta \Gamma(d+1 + \sfrac{1}{\beta})},
\quad d \geq 0  
\label{Dindist}
\end{equation}
\begin{equation}
P(\Din = d) = \frac{
\Gamma(d+1)\Gamma(1/\beta)}{\beta^2 \Gamma(d+2 + \sfrac{1}{\beta})},
\quad d \geq 0  
\label{Din=d}
\end{equation}
with asymptotics
\[
P(\Din = d) \sim \beta^{-2} \Gamma(1/\beta) \ d^{-1-\sfrac{1}{\beta}} .
\]
Formula (\ref{Din=d}) appears as a special case of recent results
in two-parameter proportional attachment models
\cite{BO03,DMS00,KRR01},
but in fact is a famous 80-year old calculation -- see section \ref{sec-Din}.

\noindent
{\bf (b).} 
The distribution of $\Dout$ is determined by the distributional equation
\begin{equation}
D \ed \left\{ \begin{array}{cc}  \sum_{i=1}^\infty \Bin(1+D_i,\alpha \lambda e^{-\lambda \xi_i})
& \low \\
 \sum_{i=1}^{\Poi(\eta)} (1+D^\prime_i) 
+ \sum_{i=1}^\infty \Bin(1+D_i,e^{-\lambda \xi_i}) & \high 
\end{array} \right.
\label{Douteq}
\end{equation}
where $D, \ D_i, D^\prime_i,\  i \geq 1$  
are i.i.d. random variables distributed as
 $\Dout$ and where
$0<\xi_1<\xi_2< \ldots$
are the points of a rate-$1$ Poisson point process on $(0,\infty)$.

We do not know how to extract a useful explicit formula from
(\ref{Douteq}) but we can compute moments.
For instance
\begin{equation}
\var \Dout = 
\left\{ \begin{array}{cc}
\frac{\alpha
(1 - \alpha + \alpha^2 \lambda /2)}
{(1-\alpha)^2
(1-\frac{1}{2}\alpha^2\lambda)
} 
& \low \\
\frac{(\eta + \frac{1}{2\lambda})(2 - \eta - \frac{1}{\lambda})}
{(1-\eta-\frac{1}{2\lambda})
(1-\eta-\frac{1}{\lambda})^2}
& \high
\end{array} \right.
\label{varDouteq}
\end{equation}
In the case $\lambda = 1/\alpha$ and in the limits
$\lambda \to 0, \ \lambda \to \infty$
we get explicit formulas for the distribution of $\Dout$ -- see
section \ref{sec-Dout} -- which show in particular that the tail of $\Dout$ has geometric
rather than power-law decay.

\noindent
{\bf (c).} 
\begin{equation}
\mbox{
$\Din$ and $\Dout$ are independent.
} 
\label{DDind}
\end{equation}

Because both $\Dout$ and $\Din$ can take the value $0$,
we see that $P(\Din + \Dout =0)>0$, implying that
$\GG_n$ will typically not be connected (see section \ref{sec-OLS}
for further comments).

\subsection{Densities of induced subgraphs}
\label{sec-induced-1}
One of the major advantages of the model is that, for a fixed
``small" graph $G$, one can (in principle, and often in practice)
calculate explicitly an ``asymptotic density"
$\dens_\infty (G)$ 
interpreted as the limit
\[ \lim_{n \to \infty}
\frac{\mbox{number of copies of $G$ in $\GG_n$}}{n}
= \dens_\infty (G) . \] 
Precise definitions are fussy, and are deferred to section
\ref{sec-induced-2}, which also records the explicit formulas
we have found.
Here let us point out the formula for triangles $K_3$:
\begin{equation}
 \dens_\infty(K_3) = \left\{ \begin{array}{cc} \frac{\alpha^3 \lambda}{(1-\alpha)(2 - \alpha^2 \lambda)} 
& \low \\
 \frac{(\eta + \frac{1}{\lambda})(\eta + \frac{1}{2\lambda})}
{(1-\eta - \frac{1}{\lambda})(1 - \eta - \frac{1}{2\lambda})}
& \high .
\end{array} \right.
\label{tri-mean-1}
\end{equation}
The formula above is the key ingredient in the formula for $\ccluster$.
Recall the verbal description
of $\ccluster$:
\begin{quote}
The proportion of directed $2$-paths
$v_1 \to v_2 \to v_3$ 
for which $v_1 \to v_3$ is also an edge.
\end{quote}
It is intuitively clear
(and formalized at (\ref{chi-path})
that the asymptotic density for occurrence
of directed $2$-paths
$v_1 \to v_2 \to v_3$,
if one does not look whether or not a third edge $v_1 \to v_2$ is present,
equals $\partial^2$
(because of independence of in-degree and out-degree at $v_2$).
So the verbal definition translates to
\[ \ccluster = \frac{\dens_\infty (K_3)}{\partial^2} \]
and then (\ref{tri-mean-1}) immediately gives formula (\ref{clust-form})
for $\ccluster$.

\subsection{Triangle density as a function of degree}
\label{tri-dens}
The parameter $\ccluster$ gives
an overall measure of triangle density.
A more detailed description is provided by statistics
$C(k), \ k \geq 2$ defined by
\[ C(k) = \frac{
E(\mbox{number of triangles containing a random degree-$k$ vertex})}
{{k \choose 2}} . \]
In principle the methods of this paper could be used to obtain an exact
formula for $C(k)$, but we shall be content with outlining
(section \ref{sec-Ck}) the tail property
\begin{equation}
C(k) \sim \frac{2 \beta_2}{\beta - \beta_2} \times \frac{1}{k}
\mbox{ as } k \to \infty .
\label{eq-Ck}
\end{equation}
See section \ref{sec-compare} for further comments.

\subsection{Edge-lengths}
\label{sec-EL}
Our model has a ``metric structure", in that there
is a distance 
$\dmetric(v,w)$ between any two vertices which does not involve
the realization of edges in the random graph.
So each edge $(v,w)$ of the graph has a real-valued length
$\dmetric(v,w)$, 
and so a typical edge has a random length $L$.
The probability density function for $L$ is given by the formula
\begin{equation}
f(x) = 
\frac{1-\alpha}{\alpha}   \sum_{i=0}^\infty \frac{(i+1)\Gamma(\alpha + 3) \ (-\lambda x)^i}{\Gamma(i + \alpha + 3)}, \ 
0 < x < \infty 
\quad \low .
\label{len-dens}
\end{equation}
{\tt Mathematica}
gives an equivalent expression as a sum of incomplete hypergeometric
functions.
One can readily observe that
$f(x) = \exp(-(\lambda \pm o(1))x)$
as $x \to \infty$.
In the underlying metric space, the number of vertices within
distance $x$ of a typical vertex grows as $e^x$.
So the tail
behavior of $f(x)$ suggests 
\begin{quote}
the chance that a vertex has an edge to its $k$'th
nearest neighbor should scale as $k^{-\lambda - 1}$
\end{quote}
though we have not attempted detailed calculations to 
verify this suggestion.
Note this property appears without being explicitly
built into the model.

\subsection{Other local statistics}
\label{sec-OLS}
There are further questions, concerning exact behavior in 
the $n \to \infty$ limit,
which are in principle solvable in terms of the limit 
network $\GG_\infty^*$, but where we have been unable to obtain usefully
explicit answers.
A major question concerns the {\em percolation probability}
\begin{equation}
p_{\mbox{\small{perc}}}(\alpha,\lambda)
= P(\mbox{typical vertex is in infinite connected component of $\GG_\infty^*$})
\label{pperc}
\end{equation}
By analogy with classical facts about the
Erd{\H{o}}s - R{\'e}nyi model, 
we expect that 
above the {\em percolation threshold},
that is when
$p_{\mbox{\small{perc}}}(\alpha,\lambda)
>0$, 
the random graph $\GG_n$ will have a {\em giant component} whose size
$C_n(\alpha,\lambda)$ satisfies
\[ n^{-1} EC_n(\alpha,\lambda) \to 
p_{\mbox{\small{perc}}}(\alpha,\lambda)
. \]
Unfortunately we do not see how to write
$p_{\mbox{\small{perc}}}(\alpha,\lambda)$
as a solution of any simple equation.
By studying an easier-to-analyze
{\em directed percolation} problem,
it is not hard to show (section \ref{sec-DP})
\begin{equation}
\mbox{if } 2 \beta - \beta_2 > 1 \mbox{ then } 
p_{\mbox{\small{perc}}}(\alpha,\lambda)
> 0 .
\label{dirpercbd}
\end{equation}

\subsection{Average distance}
\label{sec-AD}
In any graph, write
$\dgraph(v,w)$ for the minimal number of edges in any path
from $v$ to $w$.
The {\em diameter} $\Delta$ and the 
{\em average vertex-vertex distance} $\Lambda$ are defined by
\[ \Delta = \max_{v,w} \dgraph(v,w), 
\ \Lambda = \mbox{ave}_{v,w} \dgraph(v,w). \] 
In the context of a simple proportional attachment model it is
known \cite{BR03} that
\[ E\Delta_n, \ E\Lambda_n = \frac{(1+o(1)) \log n}{\log \log n}
\mbox{ as } n \to \infty . \]
It is natural to conjecture, but hard to prove, the same result for our model
(above the percolation threshold and restricted to the giant component).
On the other hand it seems likely that
standard techniques of abstract 
mathematical probability would be enough to show the weaker
bound
$ E\Lambda_n = O(\log n) \mbox{ as } n \to \infty $.
Such questions {\em cannot} in principle be answered completely using $\GG_\infty^*$.

\subsection{Summary of advantages and disadvantages of the model}
The previous sections convey some advantages of the model: 
\begin{itemize}
\item it has the three qualitative features desired in a complex
network model (power-law degree distribution, clustering,
small diameter);
\item 
it fits the complete possible range of mean degree (or scaling exponent) and clustering parameters;
\item it permits a broad range of explicit calculations. \end{itemize} So to be fair let
us list some disadvantages
from a modeling viewpoint.
\begin{itemize}
\item $\GG_n$ is not connected (for large $n$); cf. section \ref{sec-OLS};
\item  there is no power law for distribution of out-degree;
\item in-degree and out-degree are independent;
\item the scaling exponent for in-degree is determined by the mean degree;
one might prefer a model where these could be specified separately;
\item in the $n \to \infty$ limit not every finite graph is possible
as an induced subgraph (section \ref{sec-dens}).
\end{itemize}

\section{The model}
\label{sec-MODEL}
\subsection{Metric copying models}
\label{sec-metric}
Let us briefly outline a general modeling framework,
{\em metric copying models}.
Each vertex $v$ is a point in a metric space; that is,
there is some real-valued distance $d(v,w)$ between 
any two vertices $v,w$.
Given some rule for 
the positions of successive vertices 
$1,2,\ldots$,
and given a function
$p:[0,\infty) \to [0,1]$, 
we can construct random directed graphs $\GG_n$
inductively on $n$ as follows.
When vertex $n$ arrives, then\\
(i) for each directed edge $(i,j)$ of $\GG_{n-1}$,
a ``copied" edge $(n,j)$ 
is created with probability $p(d(n,i))$;\\
(ii) for each vertex $i$ ($1\leq i < n$),
a new edge $(n,i)$
is created with probability $p(d(n,i))$;\\
(iii) the events above are independent, except that repeat
edges are censored.

\vspace{0.06in}
\noindent
Imagine $p(\cdot)$ to be rapidly decreasing.
A moment's thought shows how this model resembles
proportional attachment models.
An existing vertex $v$ with in-degree $d$ has $d+1$ opportunities
to acquire an in-edge, 
due to the next arriving vertex being close to $v$ or close
to one of the $d$ vertices with edges to $v$.

In principle one 
could study such models based on random points in $d$-dimensional space, but
within such settings it is notoriously hard to do explicit calculations
(see e.g. \cite{penrose-RGG} for different models of
random graphs based on $d$-dimensional random points),
and the choice of $d$ is arbitrary.
We will avoid both problems 
by using a well known (in other contexts)
model which is loosely interpretable as ``random points in infinite-dimensional
space".
Note that in $d$-dimensional space, the number of points within
distance $r$ of a typical point grows as $r^d$; what will make our
model ``infinite-dimensional" is that this number grows as $e^r$.

For later use recall that 
a {\em pre-metric} $\bar{d}(i,j)$ is symmetric and strictly positive
for $j \neq i$.
A pre-metric can be used to
specify a metric $d(v,w)$ as the minimum, over
paths $v = i_0,i_1,\ldots,i_k = w$, of
$\bar{d}(i_0,i_1) + \bar{d}(i_1,i_2) + \ldots + \bar{d}(i_{k-1},i_k)$.

\subsection{A $d$-dimensional analogy}
\label{sec-aPP}
As a final preliminary, the following analogy may be helpful.
In $d$-dimensional space $R^d$,
take a cube $[-n^{1/d}/2,n^{1/d}/2]^d$
of volume $n$,
and put $n$ uniform random points in that cube.
This structure has a $n \to \infty$ limit,
the 
{\em Poisson point process} in $R^d$
with mean intensity $1$ point per unit volume.
Moreover the limit process, 
which is a spatial point process on all of $R^d$,
can be
represented as the distribution, at any fixed time,
of a time-evolving process of points on all of $R^d$,
where the evolution rules are\\
(i) points move away from the origin as deterministic
motion with exponential rate $1/d$; a point at position $x$ at time $t$
will be at position $xe^{(t^\prime -t)/d}$
at times $t^\prime > t$.\\
(ii)
New points arrive throughout $R^d$ as a rate-$1$ space-time Poisson process; 
that is, the chance of a point arriving in a cube of volume
$dx$ during a time interval $dt$ equals $1 \cdot dx\ dt$.\\
Thus if one takes a volume-$1$ region of space at time $t_0$,
this space expands to become volume $e^t$ at time $t_0 +t$,
and the arrival rate per unit time within this expanding
volume is $e^t$ at time $t_0 + t$.

In the limit process, one may regard the ``present time" as time $0$,
and regard the process as having evolved\footnote{This model is reminiscent of the steady-state theory of the Universe advocated by Fred Hoyle in the 1950s} 
over time
$- \infty < t \leq 0$.
Particles at the present time have ages which are independent
$\Exp(1)$ random variables independent of present positions;
from the present configuration of positions and ages one can deterministically
reconstruct the past evolution of the process.

\subsection{The stochastic mean-field model of distance}
\label{sec-SMFM}
Our model of an underlying metric space is specified by three rules.\\
(i) Point $n$ arrives at time $t_n = \log n$.\\
(ii) At the arrival time $t_n$, define the pre-distances
$(\bar{D}(n,j;t_n), \ 1 \leq j \leq n-1)$
from $n$ to the earlier-arriving points
to be independent random variables with exponential, mean $n$,
distribution.\\
(iii) Distances grow exponentially with time;
$\bar{D}(n,j;t) = e^{t - t_n}\bar{D}(n,j;t_n)$ 
for $t > t_n$.
\\
So at time $t$ there are $n = \lfloor e^t \rfloor$ points, and the
${n \choose 2}$ pre-distances
$\bar{D}(i,j;t), \ 1 \leq i < j \leq n)$
are independent random variables with exponential, mean $e^t$,
distribution.
These particular pre-distances are an instance of a pre-metric, and
this pre-metric specifies a metric
$D(i,j;t), \ 1 \leq i < j \leq n)$.
Write $(\DD_t, 0 \leq t < \infty)$ for this process of arriving points
and distances.

Here is a key feature of this construction.
At time $t$ pick a uniform random point $V_t$ as a ``root".
Then there is a $t \to \infty$ limit (in distribution)
structure, which is a metric space on a countable infinite
number of points, one being distinguished as the root.
The limit structure, called the PWIT,
is described below.
The meaning of ``limit" is that, for arbitrary fixed $r<\infty$,
the configuration of points in $\DD_t$ within distance $r$ of $V_t$
converges in distribution to the configuration 
of points of the PWIT within distance $r$ of the root
(this is {\em local weak convergence} of random networks \cite{me101}).

\subsection{The PWIT}
The PWIT is defined by a construction, illustrated in
Figure 1\footnote{Our figures are illustrations of the definitions,
rather than honest Monte Carlo simulations}.
Start with a single root vertex $\emptyset$.  This root 
vertex is then given  an \emph{infinite} number of {\em \nn s}, and the
edges from the root to the \nn s are
assigned lengths according to a realization of a Poisson process
$(\xi_i^{\emptyset}:  1 \leq i < \infty)$ of rate $1$ on $(0,\infty)$.
Now, recursively,
each vertex $v$ 
arising as a \nn\ of a previous vertex is given an infinite number
of \nn s, and the edges to these \nn s of $v$ are again assigned lengths
according to an independent realization of a Poisson process
$(\xi^v_i:  1 \leq i < \infty)$ of rate $1$.
This procedure is then continued \emph{ad infinitum}.
The resulting rooted infinite tree is a well defined random object,
called the
\emph{Poisson weighted
infinite tree} (PWIT). 

\setlength{\unitlength}{0.36in}
\begin{picture}(12,12)
\thinlines
\put(0.5,0.5){\line(1,0){11}}
\put(0.5,0.5){\line(0,1){11}}
\put(11.5,11.5){\line(0,-1){11}}
\put(11.5,11.5){\line(-1,0){11}}
\put(6,6){\circle*{0.2}}
\put(6,5){\circle*{0.2}}
\put(6,5.8){\line(0,-1){0.6}}
\put(6.1,5.3){$\xi^{\emptyset}_1$}
\put(4,8){\circle*{0.2}}
\put(5.8,6.2){\line(-1,1){1.6}}
\put(4.7,6.6){$\xi^{\emptyset}_3$}
\put(8,7){\circle*{0.2}}
\put(6.2,6.1){\line(2,1){1.6}}
\put(7.0,6.3){$\xi^{\emptyset}_2$}
\put(11,5){\circle*{0.2}}
\put(6.2,5.96){\line(5,-1){4.6}}
\put(8.5,5.1){$\xi^{\emptyset}_4$}
\put(5,4){\circle*{0.2}}
\put(5.8,4.8){\line(-1,-1){0.6}}
\put(7,2){\circle*{0.2}}
\put(6.07,4.8){\line(1,-3){0.86}}
\put(10,1){\circle*{0.2}}
\put(6.2,4.8){\line(1,-1){3.6}}
\put(4,2){\circle*{0.2}}
\put(4.9,3.8){\line(-1,-2){0.8}}
\put(5,1){\circle*{0.2}}
\put(5,3.8){\line(0,-1){2.6}}
\put(3,8){\circle*{0.2}}
\put(3.8,8){\line(-1,0){0.6}}
\put(1,10){\circle*{0.2}}
\put(2.8,8.2){\line(-1,1){1.6}}
\put(4,10){\circle*{0.2}}
\put(4,8.2){\line(0,1){1.6}}
\put(7,11){\circle*{0.2}}
\put(4.2,8.2){\line(1,1){2.6}}
\put(2,4){\circle*{0.2}}
\put(3.9,7.8){\line(-1,-2){1.8}}
\put(2,2){\circle*{0.2}}
\put(2,3.8){\line(0,-1){1.6}}
\put(9,10){\circle*{0.2}}
\put(8.07,7.2){\line(1,3){0.86}}
\put(10,8){\circle*{0.2}}
\put(8.2,7.1){\line(2,1){1.6}}
\put(10,9){\circle*{0.2}}
\put(10,8.2){\line(0,1){0.6}}
\put(11,11){\circle*{0.2}}
\put(10.07,8.2){\line(1,3){0.86}}
\put(4.17,7.94){$a$}
\put(1.66,3.93){$b$}
\put(10.07,7.69){$c$}
\put(5.6,5.9){$\emptyset$}
\end{picture}

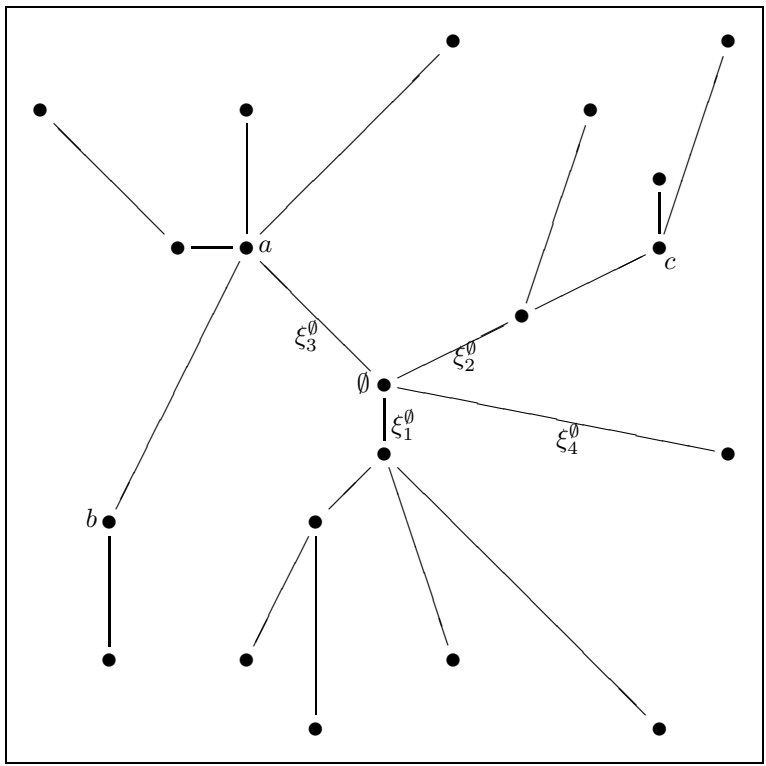
{\bf Figure 1.  The PWIT.}
{\small 
Illustration of the vertices of the PWIT within a window
of radius $3$ centered on the root $\emptyset$.
Lines indicate the \nn\ relationship, but are drawn only
when both end-vertices are within the window.
Thus the four \nn s of $\emptyset$ shown are at distances
$0< \xi^{\emptyset}_1 < \xi^{\emptyset}_2 < \xi^{\emptyset}_3 < \xi^{\emptyset}_4 < 3$
from $\emptyset$, while there are an infinite number of
\nn s of $\emptyset$ at distances greater than $3$.
Orientation of lines in pictures is arbitrary.
Labels $a,b,c$ are included for later comparisons.
}

\vspace{0.12in}
\noindent
The {\em distance} $D(v,w)$ between two vertices of the PWIT
is just the sum of edge-lengths along the path from $v$ to $w$.
Though we have drawn a tree in Figure 1,  
the lines merely indicate the \nn\ relationships;
it is better to think of the edges as absent
while retaining the distances $D(v,w)$.
In this way we may regard the vertices of the PWIT as an infinite-dimensional analog of
the $d$-dimensional Poisson point process in section \ref{sec-aPP}.
Formula (\ref{Yule-PWIT}) later provides one formalization
of ``infinite-dimensional".

The survey \cite{me101}
gives a careful explanation of how the PWIT arises as
a limit of finite models such as $\DD_t$,
and gives some applications to combinatorial optimization\footnote{See also 
\cite{me103} for novel scaling exponents arising in the
study of the mean-field traveling salesman problem.}.
The key point is that, for an arriving vertex $V_t$ in $\DD_t$,
the existing vertices at smallest $\bar{D}$-distances
correspond in the limit to the \nn s in the PWIT.
(Recall the section \ref{sec-SMFM} construction;
we are repeating the ``key feature" from the last paragraph of
that section.)

In the present setting, each point $v$ of $\DD_t$ 
has an ``age" at time $t$,
and in the limit PWIT these ages are
(exactly as in section \ref{sec-aPP})
independent $\Exp(1)$ random variables,
$A_v$ say.
Thus if we write $\DD_0^*$ for the PWIT
and $A_{\root}$ for the age of the root, then
(given the other ages $A_v$ also)
we can reconstruct the time-evolution of 
a {\em backwards space-time PWIT process}
$(\DD^*_s, \ - A_{\root} \leq s \leq 0)$.
Precisely, as $s$ runs backwards\\
(a) the edge-lengths $\xi$ decrease exponentially; at time $s<0$
the length is $\xi e^s$;\\
(b) a vertex $v$ and its incident edges are deleted at $s = -A_v$.\\
Then $\DD^*_s$ is defined as the connected component containing the root at time $s$.
This limit process relates to the finite process as follows.
Let $\tilde{A}_t$ be the age (at time $t$) of the randomly-chosen vertex $V_t$
at time $t$.  Then
\begin{equation}
 (\DD^{*(t)}_{t+s}, \ - \tilde{A}_t \leq s \leq 0)
\cd 
(\DD^*_s, \ - \tilde{A}_{\root} \leq s \leq 0)
\mbox{ as } t \to \infty \label{D*1}
\end{equation}
where $\DD^{*(t)}_{t+s}$ is the configuration $\DD_{t+s}$ rooted at $V_t$.

There is also a
{\em forwards space-time PWIT process}
$(\DD^*_s, 0 \leq s < \infty)$
specified as follows.
Start with the PWIT $\DD_0^*$.
At time $s$ increases, all inter-vertex distances
increase at exponential rate $1$.
For each vertex $v$ present at time $s$, and each $0<r<\infty$,
there is (as explained below) chance $1 \cdot dr \ ds$ that during
$[s,s+ds]$ a new vertex $v^\prime$ will appear at distance 
$\in [r,r+dr]$ from $v$ as a \nn\ of $v$.
Along with this vertex 
(which has current age $0$)
is an independent copy of the PWIT rooted at $v^\prime$,
whose other vertex-ages are independent $\Exp(1)$.
The relation between the finite-$t$ and the limit process
is analogous to (\ref{D*1}):
\begin{equation}
 (\DD^{*(t)}_{t+s}, \ 0 \leq s < \infty)
\cd 
(\DD^*_s, \ 0 \leq s < \infty)
\mbox{ as } t \to \infty . \label{D*2}
\end{equation}
Here is the calculation leading to the coefficient ``$1$" in
\begin{eqnarray}
\mbox{there is chance $1 \cdot dr \ ds$ that during
$[s,s+ds]$ a new vertex $v^\prime$ }&&\nonumber\\ \mbox{will appear at distance 
$\in [r,r+dr]$ from $v$ as a \nn\ of $v$
} . \label{newvertex}
\end{eqnarray}
In the process $(\DD_t)$, during time $[t+s,t+s+ds]$
about $e^{t+s}ds$ vertices arrive;
for each existing vertex, the chance an arriving vertex is
within $\bar{D}$-distance $[r,r+dr]$ equals
$e^{-(t+s)} \exp(-re^{-(t+s)})dr$.
So the chance in (\ref{newvertex}) equals
\[ e^{t+s}ds \times 
e^{-(t+s)} \exp(-re^{-(t+s)})dr
\approx 1 \times dr \ ds . \]

\paragraph{Recursive self-similarity.}
Implicit in the model is the fact that the ``geometry" of the
space seen by a newly-arriving particle $v^*$ is statistically the
same as the geometry seen by a typical existing particle.
This is the familiar PASTA
(Poisson arrivals see time averages)
property in queuing theory.
In particular, at the arrival time of $v^*$ the
{\em geometric components}
containing the different \nn\ vertices $v_1,v_2,\ldots$
are independent copies of the PWIT.  This
{\em recursive self-similarity} property of the PWIT process
is fundamental to its analytic tractability.

Figure 2 and its legend may be helpful.

\setlength{\unitlength}{0.35in}
\begin{picture}(16,7)(1,0)
\put(0,0){\line(1,0){6}}
\put(0,0){\line(0,1){6}}
\put(6,6){\line(0,-1){6}}
\put(6,6){\line(-1,0){6}}
\put(3,3){\circle*{0.2}}
\put(2.5,2){\circle*{0.2}}
\put(2.9,2.8){\line(-1,-2){0.3}}
\put(3,0.5){\circle*{0.2}}
\put(2.57,1.8){\line(1,-3){0.36}}
\put(2,5){\circle*{0.2}}
\put(2.9,3.2){\line(-1,2){0.8}}
\put(1,4){\circle*{0.2}}
\put(1.8,4.8){\line(-1,-1){0.6}}
\put(5.5,4.25){\circle*{0.2}}
\put(3.2,3.1){\line(2,1){2.1}}
\put(9,0){\line(1,0){6}}
\put(9,0){\line(0,1){6}}
\put(15,6){\line(0,-1){6}}
\put(15,6){\line(-1,0){6}}
\put(12,3){\circle*{0.2}}
\put(10,4){\circle*{0.2}}
\put(11.8,3.1){\line(-2,1){1.6}}
\put(14,5){\circle*{0.2}}
\put(12.2,3.2){\line(1,1){1.6}}
\put(13,2){\circle*{0.2}}
\put(12.2,2.8){\line(1,-1){0.6}}
\put(13,0.5){\circle*{0.2}}
\put(13,1.8){\line(0,-1){1.1}}
\put(14.5,2){\circle*{0.2}}
\put(13.2,2){\line(1,0){1.1}}
\put(11,5){\circle*{0.2}}
\put(10.2,4.2){\line(1,1){0.6}}
\put(3,3.21){$a$}
\put(2.12,1.95){$b$}
\put(11.7,2.7){$c$}
\end{picture}

\vspace{0.16in}
\noindent
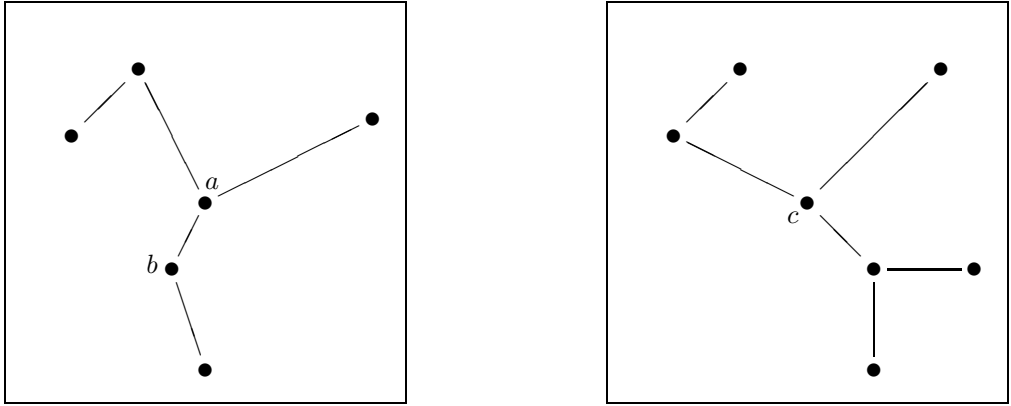
{\bf Figure 2.  The space-time PWIT process.}
{\small 
Regarding Figure 1 as showing the PWIT at a time $t_+$, 
Figure 2 shows the space-time PWIT at an earlier time $t_-$
at which only three vertices $a,b,c$ of the vertices in the
Figure 1 window have arrived.  
Figure 2 shows smaller windows centered on $a$ and on $c$.
The other vertices in Figure 2, and the \nn\ relation shown by lines,
are still present at time $t_+$,
but are not visible in Figure 1 because the expansion of distances
has placed them outside the Figure 1 window.
}

\subsection{The MFSC model}
\label{sec-MFSC}
The process
$(\DD_{t_n}, n = 1,2,3,\ldots)$
of arrivals and inter-point distances described in
section \ref{sec-SMFM}
defines an ``underlying geometry";
we now define the random graph process
$(\GG_n, n = 1,2,3,\ldots)$
which is the subject of this paper.
Fix two parameters
$0<\alpha < \infty$
and
$0<\lambda<\infty$.
Write
\begin{equation}
 p(x) = \min(1,\alpha \lambda e^{-\lambda x}), \quad 0 \leq x < \infty . \label{pdef}
\end{equation}
We now implement a version of the ``metric copying" idea from section \ref{sec-metric}.
$\GG_1$ consists of vertex $1$ and no edges.
When vertex $n$ arrives at time $t_n = \log n$, then\\
(i) for each directed edge $(i,j)$ of $\GG_{n-1}$,
a ``copied" edge $(n,j)$ 
is created with probability $p(\bar{D}(n,i;t_n))$;\\
(ii) for each vertex $i$ ($1\leq i < n$),
a new edge $(n,i)$
is created with probability $p(\bar{D}(n,i;t_n))$;\\
(iii) the events above are independent, except that repeat
edges are censored.

\vspace{0.08in}
\noindent
Note that we use $\bar{D}$ instead of $D$ in determining
attachment probabilities
(because $\bar{D}$-near vertices at finite time correspond to the \nn s 
in the limit PWIT).
We call 
$(\GG_n, n = 1,2,3,\ldots)$
the
{\em mean-field simple copying} (MFSC)
model.

\setlength{\unitlength}{0.35in}
\begin{picture}(16,6.3)(1,1)
\put(9,0){\line(1,0){6}}
\put(9,0){\line(0,1){6}}
\put(15,6){\line(0,-1){6}}
\put(15,6){\line(-1,0){6}}
\thicklines
\put(12,3){\circle*{0.2}}
\put(10,4){\circle*{0.2}}
\put(14,5){\circle*{0.2}}
\put(12.2,3.2){\vector(1,1){1.6}}
\put(15.8,5){\vector(-1,0){1.6}}
\put(14,5.2){\vector(0,1){1.6}}
\put(13,2){\circle*{0.2}}
\put(12,2.8){\vector(0,-1){3.6}}
\put(12.94,1.8){\vector(-1,-3){0.88}}
\put(13.94,-0.8){\vector(-1,3){0.88}}
\put(12.2,3){\vector(1,0){4.1}}
\put(13.3,2.16){\vector(4,1){3.0}}
\put(13.2,0.6){\vector(3,2){3.1}}
\put(13,0.5){\circle*{0.2}}
\put(13,0.7){\vector(0,1){1.1}}
\put(14.5,2){\circle*{0.2}}
\put(14.3,2){\vector(-1,0){1.1}}
\put(11,5){\circle*{0.2}}
\put(10.8,4.8){\vector(-1,-1){0.6}}
\put(9.8,4){\vector(-1,0){1.6}}
\put(10.8,4.94){\vector(-3,-1){2.6}}
\put(9.8,4.2){\vector(-1,1){1.6}}
\put(8.2,2.2){\vector(1,1){1.6}}
\thinlines
\put(0,0){\line(1,0){6}}
\put(0,0){\line(0,1){6}}
\put(6,6){\line(0,-1){6}}
\put(6,6){\line(-1,0){6}}
\thicklines
\put(3,3){\circle*{0.2}}
\put(2.5,2){\circle*{0.2}}
\put(2.9,2.8){\vector(-1,-2){0.3}}
\put(3,0.5){\circle*{0.2}}
\put(2.93,0.7){\vector(-1,3){0.36}}
\put(3,0.3){\vector(-3,-2){1.1}}
\put(2.4,1.8){\vector(-1,-3){0.8}}
\put(5.3,-0.8){\vector(-1,1){2.6}}
\put(2,5){\circle*{0.2}}
\put(2.8,3.1){\vector(-2,1){1.6}}
\put(1,4){\circle*{0.2}}
\put(1.8,4.8){\vector(-1,-1){0.6}}
\put(1.8,5){\vector(-1,0){2.6}}
\put(1,6.8){\vector(0,-1){2.6}}
\put(0.8,4.1){\vector(-2,1){1.6}}
\put(-0.8,4){\vector(1,0){1.6}}
\put(0.8,3.8){\vector(-1,-1){1.6}}
\put(5.5,4.25){\circle*{0.2}}
\put(5.3,4.15){\vector(-2,-1){2.1}}
\put(3,3.21){$a$}
\put(2.12,1.95){$b$}
\put(11.7,2.7){$c$}
\end{picture}

\vspace{0.7in}
\noindent
{\bf Figure 3.}
{\small 
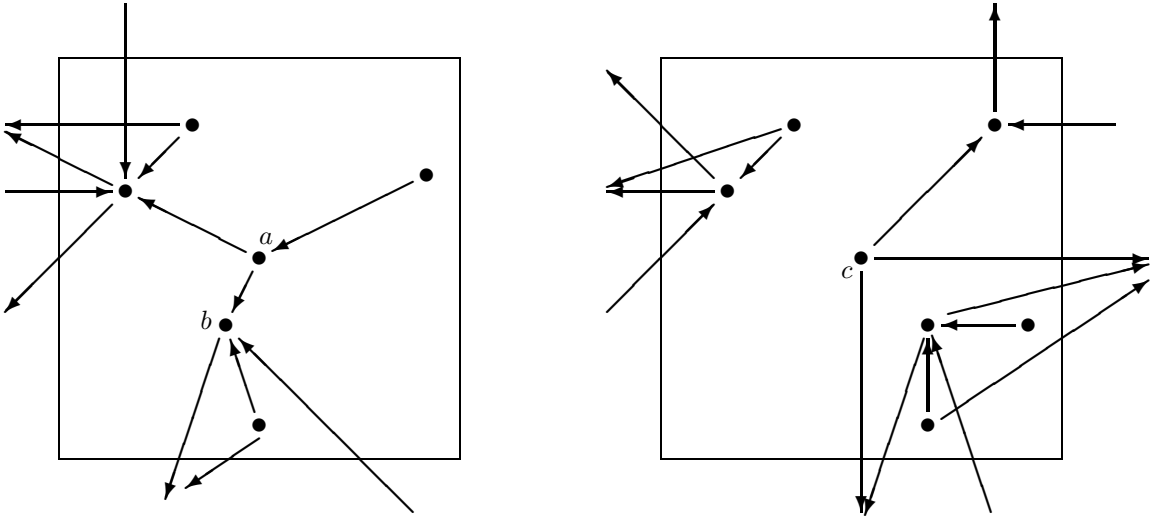
The graph process $\GG^*_\infty(t_-)$ on the realization
of the space-time PWIT at time $t_-$ in Figure 2.
For the graph process we show all edges with either end-vertex within
the window.
The following figures show the evolution of $\GG^*_\infty(t)$
over $t_- < t \leq t_+$.
}

\newpage
\setlength{\unitlength}{0.25in}
\begin{picture}(12,8.6)
\thinlines
\put(0.5,0.5){\line(1,0){11}}
\put(0.5,0.5){\line(0,1){11}}
\put(11.5,11.5){\line(0,-1){11}}
\put(11.5,11.5){\line(-1,0){11}}
\thicklines
\put(4,8){\circle*{0.2}}
\put(4.17,7.94){$a$}
\put(2,4){\circle*{0.2}}
\put(2.17,3.91){$b$}
\put(3.8,7.8){\vector(-1,-2){1.8}}
\put(1.8,4){\vector(-1,0){2.6}}
\put(-0.8,2.93){\vector(3,1){2.6}}
\put(-0.8,5.13){\vector(3,-1){2.6}}
\put(5.25,13){\vector(-1,-4){1.2}}
\put(3.9,7.97){\vector(-4,-1){4.9}}
\put(10,8){\circle*{0.2}}
\put(9.66,7.93){$c$}
\put(10.2,8){\vector(1,0){2.6}}
\put(10.2,8.07){\vector(3,1){2.6}}
\put(10.2,7.93){\vector(3,-1){2.6}}
\end{picture}

{\bf Figure 4.}
{\small 
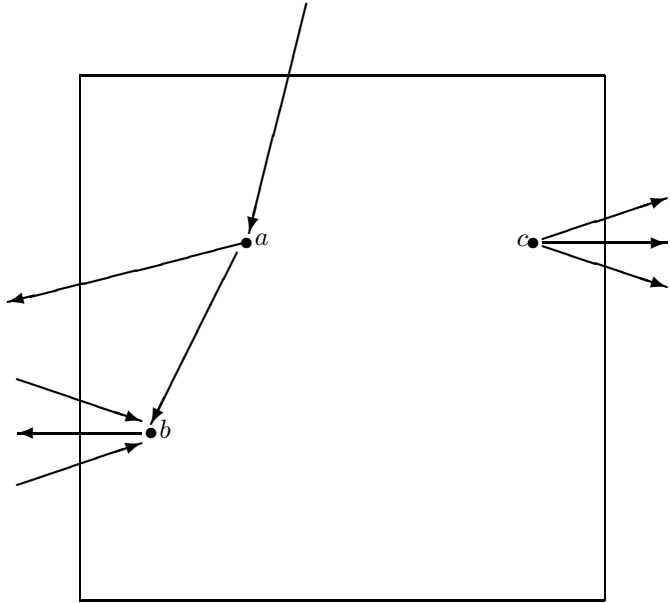
Figures 4 -- 6 build up the graph $\GG^*_\infty(t_+)$
on the time-$t_+$ PWIT in Figure 1.
Figure 4 here shows only the edges that were present at time
$t_-$, that is the edges shown in Figure 3.
Some edges crossing outside the window have been redrawn at
different angles for later convenience.
}

\setlength{\unitlength}{0.25in}
\begin{picture}(12,12)(0,2)
\thinlines
\put(0.5,0.5){\line(1,0){11}}
\put(0.5,0.5){\line(0,1){11}}
\put(11.5,11.5){\line(0,-1){11}}
\put(11.5,11.5){\line(-1,0){11}}
\thicklines
\put(1,10){\circle*{0.2}}
\put(0.8,9.93){\vector(-3,-1){1.6}}
\put(0.8,10.07){\vector(-3,1){1.6}}
\put(1,12.8){\vector(0,-1){2.6}}
\put(7,11){\circle*{0.2}}
\put(7,11.2){\vector(0,1){1.6}}
\put(8.8,12.8){\vector(-1,-1){1.6}}
\put(4,10){\circle*{0.2}}
\put(4,9.8){\vector(0,-1){1.6}}
\put(3.93,9.8){\vector(-1,-3){1.86}}
\put(3,8){\circle*{0.2}}
\put(3.2,8){\vector(1,0){0.6}}
\put(2.8,7.93){\vector(-3,-1){3.6}}
\put(2,2){\circle*{0.2}}
\put(2,2.2){\vector(0,1){1.6}}
\put(1.07,-0.8){\vector(1,3){0.86}}
\put(2.93,-0.8){\vector(-1,3){0.86}}
\put(4,2){\circle*{0.2}}
\put(4.9,3.8){\vector(-1,-2){0.8}}
\put(4,-0.8){\vector(0,1){2.6}}
\put(4.8,1.2){\vector(-1,1){0.6}}
\put(5,1){\circle*{0.2}}
\put(5,1.2){\vector(0,1){2.6}}
\put(5.2,1.1){\vector(2,1){1.6}}
\put(5,4){\circle*{0.2}}
\put(5.2,4.2){\vector(1,1){0.6}}
\put(5.2,3.8){\vector(1,-1){1.6}}
\put(6,5){\circle*{0.2}}
\put(6.07,4.8){\vector(1,-3){0.86}}
\put(7,-0.8){\vector(0,1){2.6}}
\put(7.93,-0.8){\vector(-1,3){0.86}}
\put(10,1){\circle*{0.2}}
\put(10,0.8){\vector(0,-1){1.6}}
\put(12.8,1){\vector(-1,0){2.6}}
\put(11,5){\circle*{0.2}}
\put(11.2,5){\vector(1,0){1.6}}
\put(12.8,4.1){\vector(-2,1){1.6}}
\put(12.8,3.2){\vector(-1,1){1.6}}
\put(9,10){\circle*{0.2}}
\put(8.93,9.8){\vector(-1,-3){0.86}}
\put(8,7){\circle*{0.2}}
\put(8.2,7.1){\vector(2,1){1.6}}
\put(8.2,7){\vector(4,-1){3.7}}
\put(10,9){\circle*{0.2}}
\put(10,8.8){\vector(0,-1){0.6}}
\put(10.2,8.93){\vector(3,-1){2.6}}
\put(11,11){\circle*{0.2}}
\put(10.93,10.8){\vector(-1,-3){0.86}}
\put(11.2,10.8){\vector(2,-3){1.6}}
\put(7,2){\circle*{0.2}}
\thinlines
\put(0.5,0.5){\line(1,0){11}}
\put(0.5,0.5){\line(0,1){11}}
\put(11.5,11.5){\line(0,-1){11}}
\put(11.5,11.5){\line(-1,0){11}}
\thicklines
\put(4,8){\circle*{0.2}}
\put(4.17,7.94){$a$}
\put(2,4){\circle*{0.2}}
\put(2.17,3.91){$b$}
\put(3.8,7.8){\vector(-1,-2){1.8}}
\put(1.8,4){\vector(-1,0){2.6}}
\put(-0.8,2.93){\vector(3,1){2.6}}
\put(-0.8,5.13){\vector(3,-1){2.6}}
\put(5.25,13){\vector(-1,-4){1.2}}
\put(3.9,7.97){\vector(-4,-1){4.9}}
\put(10,8){\circle*{0.2}}
\put(9.66,7.93){$c$}
\put(10.2,8){\vector(1,0){2.6}}
\put(10.2,8.07){\vector(3,1){2.6}}
\put(10.2,7.93){\vector(3,-1){2.6}}
\end{picture}

\vspace{0.7in}
\noindent
{\bf Figure 5.}
{\small 
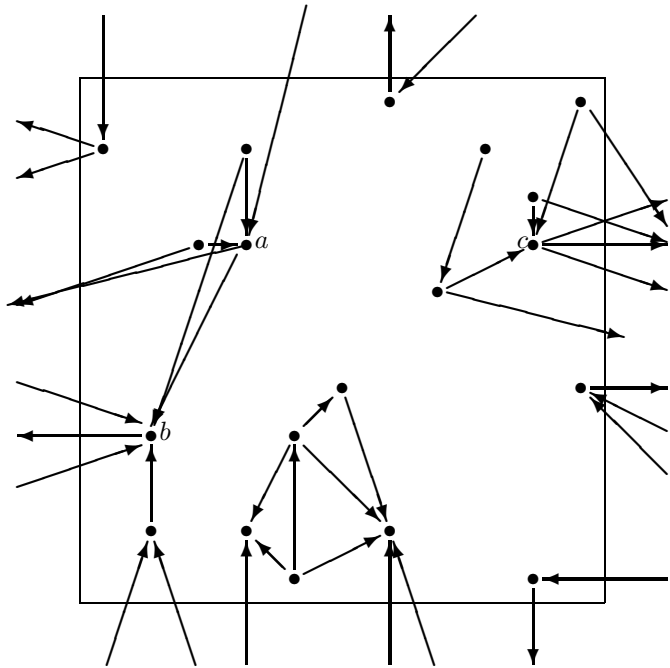
Suppose in Figure 1 that the center vertex $\emptyset$ arrives
at time $t_+$; Figure 5 illustrates the graph just before that arrival.
Since the time-$t_-$ configuration in Figure 3,
more vertices have arrived and formed edges, and 
distances have expanded.
}

\newpage
\setlength{\unitlength}{0.25in}
\begin{picture}(12,12)
\thinlines
\put(0.5,0.5){\line(1,0){11}}
\put(0.5,0.5){\line(0,1){11}}
\put(11.5,11.5){\line(0,-1){11}}
\put(11.5,11.5){\line(-1,0){11}}
\thicklines
\put(1,10){\circle*{0.2}}
\put(0.8,9.93){\vector(-3,-1){1.6}}
\put(0.8,10.07){\vector(-3,1){1.6}}
\put(1,12.8){\vector(0,-1){2.6}}
\put(7,11){\circle*{0.2}}
\put(7,11.2){\vector(0,1){1.6}}
\put(8.8,12.8){\vector(-1,-1){1.6}}
\put(4,10){\circle*{0.2}}
\put(4,9.8){\vector(0,-1){1.6}}
\put(3.93,9.8){\vector(-1,-3){1.86}}
\put(3,8){\circle*{0.2}}
\put(3.2,8){\vector(1,0){0.6}}
\put(2.8,7.93){\vector(-3,-1){3.6}}
\put(2,2){\circle*{0.2}}
\put(2,2.2){\vector(0,1){1.6}}
\put(1.07,-0.8){\vector(1,3){0.86}}
\put(2.93,-0.8){\vector(-1,3){0.86}}
\put(4,2){\circle*{0.2}}
\put(4.9,3.8){\vector(-1,-2){0.8}}
\put(4,-0.8){\vector(0,1){2.6}}
\put(4.8,1.2){\vector(-1,1){0.6}}
\put(5,1){\circle*{0.2}}
\put(5,1.2){\vector(0,1){2.6}}
\put(5.2,1.1){\vector(2,1){1.6}}
\put(5,4){\circle*{0.2}}
\put(5.2,4.2){\vector(1,1){0.6}}
\put(5.2,3.8){\vector(1,-1){1.6}}
\put(6,5){\circle*{0.2}}
\put(6.07,4.8){\vector(1,-3){0.86}}
\put(7,-0.8){\vector(0,1){2.6}}
\put(7.93,-0.8){\vector(-1,3){0.86}}
\put(10,1){\circle*{0.2}}
\put(10,0.8){\vector(0,-1){1.6}}
\put(12.8,1){\vector(-1,0){2.6}}
\put(11,5){\circle*{0.2}}
\put(11.2,5){\vector(1,0){1.6}}
\put(12.8,4.1){\vector(-2,1){1.6}}
\put(12.8,3.2){\vector(-1,1){1.6}}
\put(9,10){\circle*{0.2}}
\put(8.93,9.8){\vector(-1,-3){0.86}}
\put(8,7){\circle*{0.2}}
\put(8.2,7.1){\vector(2,1){1.6}}
\put(8.2,7){\vector(4,-1){3.7}}
\put(10,9){\circle*{0.2}}
\put(10,8.8){\vector(0,-1){0.6}}
\put(10.2,8.93){\vector(3,-1){2.6}}
\put(11,11){\circle*{0.2}}
\put(10.93,10.8){\vector(-1,-3){0.86}}
\put(11.2,10.8){\vector(2,-3){1.6}}
\put(7,2){\circle*{0.2}}
\thinlines
\put(0.5,0.5){\line(1,0){11}}
\put(0.5,0.5){\line(0,1){11}}
\put(11.5,11.5){\line(0,-1){11}}
\put(11.5,11.5){\line(-1,0){11}}
\thicklines
\put(4,8){\circle*{0.2}}
\put(4.17,7.94){$a$}
\put(2,4){\circle*{0.2}}
\put(2.17,3.91){$b$}
\put(3.8,7.8){\vector(-1,-2){1.8}}
\put(1.8,4){\vector(-1,0){2.6}}
\put(-0.8,2.93){\vector(3,1){2.6}}
\put(-0.8,5.13){\vector(3,-1){2.6}}
\put(5.25,13){\vector(-1,-4){1.2}}
\put(3.9,7.97){\vector(-4,-1){4.9}}
\put(10,8){\circle*{0.2}}
\put(9.66,7.93){$c$}
\put(10.2,8){\vector(1,0){2.6}}
\put(10.2,8.07){\vector(3,1){2.6}}
\put(10.2,7.93){\vector(3,-1){2.6}}
\put(5.96,5.8){\vector(0,-1){0.6}}
\put(6.09,5.8){\vector(1,-4){0.9}}
\put(5.8,6.2){\vector(-1,1){1.6}}
\put(6.2,6){\vector(1,0){5.7}}
\put(6,6){\circle*{0.2}}
\put(5.66,4.93){$d$}
\put(5.57,5.93){$\emptyset$}
\put(7.66,6.93){$e$}
\put(10.66,4.93){$f$}
\end{picture}

\vspace{0.4in}
\noindent
{\bf Figure 6.}
{\small 
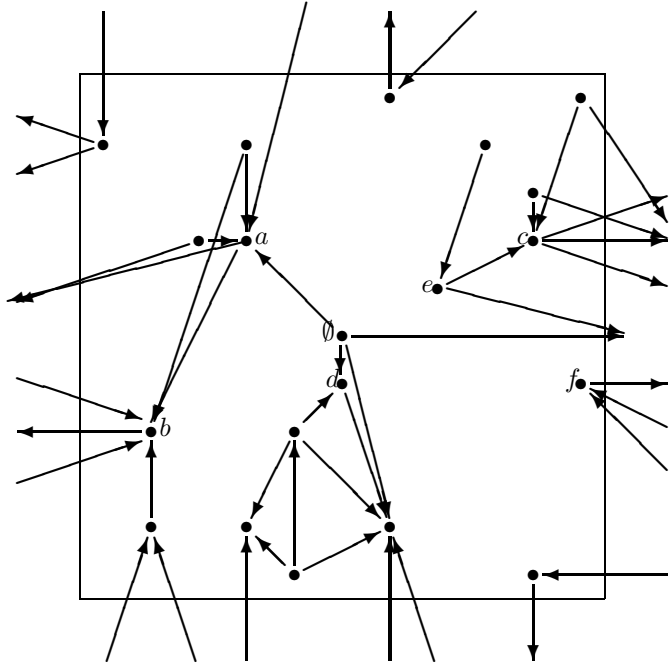
Vertex $\emptyset$ arrives at time $t_+$ with \nn s 
$d,e,a,f,\ldots$.
Out-edges from $\emptyset$ appear according to the metric copying
scheme of section \ref{sec-metric} with $p(x)$ given by (\ref{pdef}).
So for each near neighbor $v$ and each existing edge $(v,w)$,
an edge $(\emptyset,v)$ or $(\emptyset,w)$ appears
with probability $p(D(\emptyset,v))$.
In this realization, edges appeared to the \nn s $d$ and $a$,
and two other edges (one from $d$ and one from $e$) were copied.
}

\vspace{0.1in}
\noindent
Our focus in this paper is the study of the MFSC model using its 
limit structure.
Just as the $t \to \infty$ limit of the time-$t$ ``geometry" $\DD_t$
is the PWIT, we can consider $\GG_t$ as a structure built over $\DD_t$,
and we get a limit random directed graph $\GG_\infty^*$ as a structure built over
the PWIT.
The structure of $\GG_\infty^*$ near the root is exactly the $t \to \infty$
limit structure of $\GG_t$ relative to a random (``typical") vertex $V_t$,
and so we can obtain (in principle) a wide variety of asymptotic 
results about $\GG_t$ by doing calculations upon $\GG_\infty^*$.
The only difficulty in this methodology is that we don't have a useful explicit
description of $\GG_\infty^*$.  
Instead, we can consider the space-time limits
(\ref{D*1},\ref{D*2}) jointly with $(\GG_t)$; then in the limit we get
the space-time PWIT processes existing jointly with a random graph
process
$(\GG_\infty^*(s))$,
where now $\GG_\infty^*(0) = \GG_\infty^*$.
The process $(\GG_\infty^*(s))$ evolves with $s$ and the space-time PWIT process
by the rules implied by (i)-(iii) above.  

Precisely, the evolution rules (illustrated by Figures 3 -- 6) are
\begin{quote}
when a new vertex $v^*$ arrives in the
forwards space-time PWIT process
at time $s$,
it has \nn s $(v_1,v_2,\ldots)$ at distances $(\xi_1,\xi_2,\ldots)$,
and $\GG_\infty^*(s)$ has put a random graph structure on the 
geometric component
containing each $v_i$.
For each $i$ and each directed edge $(v_i,w)$, a new edge
$(v^*,v_i)$ or $(v^*,w)$ is created with probability 
$p(D(v^*,v_i))$, independently for different possible edges.
\end{quote}

\paragraph{Recursive self-similarity.}
What makes this process tractable is that the
recursive self-similarity property of the PWIT
extends to the random graph process; each $v_i$ defines a 
geometric component and a random graph on that component,
and these are independent copies of the joint distribution of
the PWIT and $\GG_\infty^*$. 
This property is used extensively in the calculations
in section \ref{sec-CALC}.

\subsection{Reparametrization and extreme cases}
\label{sec-extreme}
Although the MFSC model makes sense for the full range 
($0 < \alpha, \lambda < \infty$)
of parameters, we will only consider the ranges 
(\ref{range-low},\ref{range-high})
for which the limit mean degree is finite.

Note that definition (\ref{pdef}) of $p(x)$
can be rewritten as
\begin{eqnarray}
p(x) &=& \alpha \lambda e^{-\lambda x},
0 < x < \infty
\quad \quad \low \label{p2low}\\
p(x) &=& \left\{
\begin{array}{ll}
1&0<x\leq \eta \\
e^{-\lambda (x-\eta)} & \eta < x < \infty 
\end{array} \right.
\quad \quad \high \label{p2high}
\end{eqnarray}
where the reparametrization
$\eta := \lambda^{-1} \log (\alpha \lambda)$
when $\alpha \lambda > 1$ 
is used in (\ref{p2high}).
In the two extremes of clustering,
our model simplifies in different ways.
For fixed $\alpha$, 
when $n$ is large and $\alpha$ is small,
the model resembles
the following variant of the proportional
attachment model:
\begin{quote}
An arriving vertex has a $\Poi(\alpha)$ number of out-edges,
whose end-vertices are chosen with probabilities proportional
to $1 + $ in-degree.
\end{quote}
At the other extreme, for fixed $\eta$  
our model makes sense with $\lambda = \infty$,
interpreting (\ref{p2high}) to mean
$p(x) = 0, \ x > \eta$.
In this case the model becomes, for large $n$,
\begin{quote}
An arriving vertex $v^*$ chooses at random a $\Poi(\eta)$ number of
neighbors $v_j$, and creates edges $(v^*,v_j)$, and also 
copies each existing edge $(v_j,w)$ to a new edge $(v^*,w)$.
\end{quote}
Clearly in the former limit we have
$\ccluster = 0$
and in the latter limit we have $\ccluster = 1$.

\section{Calculations}
\label{sec-CALC}
In section \ref{sec-CALC} we derive the formulas stated in
sections \ref{sec-2par} - \ref{sec-EL}.
As described in section \ref{sec-MFSC},
our methodology is to regard $\Din$ and $\Dout$ as the (random)
in-degree and out-degree of the root in $\GG_\infty^*$,
and to study this using the time-dynamics of $\GG_\infty^*(s)$
derived from the space-time PWIT process and the evolution rules
of the graph process.

\subsection{Two helpful calculations}
We will make frequent use of the next lemma.
\begin{Lemma}
\label{LZu}
For integers $u \geq 1$ write
\[ Z^{(u)} = \sum_{i=1}^\infty p^u(\xi_i) = \sum_{i=1}^\infty
\left[ \min(1,\alpha \lambda e^{-\lambda \xi_i}) \right]^u . \]
Then
\begin{equation}
\beta_u := E Z^{(u)} =
\left\{ \begin{array}{cc}
u^{-1} \alpha^u \lambda^{u-1} & \mathrm{[low]} \\
\eta + \sfrac{1}{u \lambda} & \mathrm{[high]}
\end{array} \right.
\label{EZZ}
\end{equation}
\begin{equation}
\var Z^{(1)} =
\left\{ \begin{array}{cc}
\sfrac{1}{2}\alpha^2 \lambda & \mathrm{[low]} \\
\eta + \sfrac{1}{2 \lambda} & \mathrm{[high]} .
\end{array} \right.
\end{equation}
In particular, $\beta:= \beta_1 < 1$ for all parameter values.
\end{Lemma}
{\em Proof.}
We will do the low clustering density case -- the high density
case is similar.
By (\ref{pp-rate})
the chance that some $\xi_i$ falls into an interval
$[x,x+dx]$ is $dx$, so
\[ \beta_u
= \int_0^\infty (\alpha \lambda e^{-\lambda x})^u \ dx
= \alpha^u \lambda^u \int_0^\infty e^{-u \lambda x} \ dx
= \alpha^u \lambda^u /(u \lambda) . \]
Moreover, there is a general formula for variance of a sum
over a Poisson (rate $1$) process $(\xi_i)$:
\[ \var (\sum_i w(\xi_i)) = \int_0^\infty w^2(x) \ dx \]
and applying this formula for $Z^{(1)}$ gives
\[ \var Z^{(1)} = \int_0^\infty (\alpha \lambda e^{-\lambda x})^2 \ dx
= \alpha^2 \lambda^2 /(2 \lambda) . \]

We next recall a classical result.
Fix $0< \theta < \infty$.
Set $N(0) = 1$ and let
$(N(t), t \geq 0)$
be the
{\em Yule process} of rate $\theta$,
that is the Markov process 
which changes only by $+1$ steps and for which
\[ P(N(t+dt) = n+1|N(t) = n) = \theta n \ dt .\]
A textbook result (e.g. \cite{ros83} sec. 5.3) says
\begin{equation}
N(t) \ed \Geo(e^{-\theta t}) .
\label{Yule-1}
\end{equation}
Note that in the PWIT, if $N(r)$ is the number of vertices
within distance $r$ from the root (counting the root itself),
then the process 
$(N(r), r \geq 0)$
is a Yule process of rate $1$, 
because for a vertex $v$ at distance $r^\prime < r$,
the chance of $v$ having a \nn\ at distance 
$\in [r-r^\prime, r-r^\prime + dr]$
equals $1 \cdot dr$.
So in particular,
\begin{equation}
 EN(r) = E \Geo(e^{-r}) = e^r . \label{Yule-PWIT}
\end{equation}

\subsection{Distribution of in-degree}
\label{sec-Din}
We start by giving the derivation of 
\[
1 + \Din \ed \Geo(e^{- \beta T})
\mbox{ where } 
T \ed \Exp(1) 
\quad \quad (\ref{Dineq}) \]
for $\beta = EZ^{(1)}$.
In the 
forwards space-time PWIT process,
let $N(t)$ be $1 + $ the in-degree of the root, 
when the root has age $t$.
Thus $N(t)$ counts the set of vertices $v$ for which
$v \to \root$ is an edge, or $v = \root$.
When a new vertex $v^\prime$ arrives with some $v$ in this set
as a \nn, 
at distance $r$, there is chance $p(r)$ for the root's in-degree
to increase by $1$, 
and so from the dynamics (\ref{newvertex}) of the
forwards space-time PWIT process
we see that $N(t)$ is the Yule process of rate
\[ \beta = \int_0^\infty p(r) \ dr . \]
Use formula (\ref{Yule-1}) and the fact that the age of the root of the PWIT
has $\Exp(1)$ distribution
to obtain (\ref{Dineq}).

We can quickly use (\ref{Dineq}) to calculate $E\Din$.
\begin{eqnarray*}
1 + E\Din &=& E(E(\Geo(e^{- \beta T})|T)) \\
&=& E e^{\beta T} 
\mbox{ because } E\Geo(p) = p^{-1} \\
&=& \int_0^\infty e^{\beta t} e^{-t} \ dt 
= \frac{1}{1-\beta}
\end{eqnarray*}
giving $E \Din = \frac{\beta}{1-\beta}$
as at (\ref{EDeq}).
We now calculate the distribution of $\Din$ in the same way.
Because
$P(\Geo(p)  \geq i+1) = (1-p)^i, \ i \geq 0$
we have
\begin{eqnarray*}
P(\Din \geq i) &=& EP(\Din \geq i|T)\\
&=& E P(1+\Din \geq i+1|T)\\
&=& E \left(1 - e^{-\beta T} \right)^i \\
&=& \int_0^\infty (1 - e^{-\beta t})^i e^{-t} \ dt \\
&=&
\sfrac{1}{\beta} \int_0^1 (1-s)^i s^{\sfrac{1}{\beta}-1} \ ds
\mbox{ setting } s = e^{-\beta t}\\
&=& \frac{1}{\beta} \ 
\frac{\Gamma(i+1)\Gamma(\sfrac{1}{\beta})}
{\Gamma(i+1+\sfrac{1}{\beta})}
\mbox{ using the Beta integral formula}.
\end{eqnarray*}
This is (\ref{Dindist}),
and (\ref{Din=d}) follows.

{\em Historical note.}
Yule \cite{yule24} introduced what we now call the Yule process in 1924
in the context of
a model for
evolution of new species.
It is interesting that his central mathematical results
are the Geometric distribution (\ref{Yule-1}) [his (5)]
and the calculation 
starting from our (\ref{Dineq}) [representing, for Yule, a
distribution of numbers of species in a typical genus]
of the explicit distribution (\ref{Din=d}) [his (12)].
After 80 years we have slicker notation but the argument is the same!
Moreover Yule's motivation was to find a simple model yielding
a power-law distribution for number of species per genus,
just as the motivation for the recent literature on proportional attachment models
was to find a simple model yielding power-law
degree distributions.

\subsection{Distribution of out-degree}
\label{sec-Dout}
We will first derive (\ref{Douteq}).
Because the out-edges are formed on arrival, we may suppose
the root of the PWIT has just arrived.
Consider a \nn\ $v^\prime$ at distance $r$.
For each out-edge of $v^\prime$,
and for $v^\prime$ itself, there is chance $p(r)$ that
a corresponding out-edge is created at the root,
giving a total number 
$\Bin(1+D(v^\prime),p(r))$ of out-edges,
where $D(v^\prime)$ is the out-degree of $v^\prime$.
The recursive self-similarity property
(end of section \ref{sec-MFSC})
implies that the
$(D(v^\prime): \ v^\prime \mbox{ \nn\ of root})$
are i.i.d. random variables distributed as $\Dout$,
and independent of their distances $(\xi_i)$ from the root.
Rewriting $(D(v^\prime))$ as $(D^{(i)})$ in increasing order
of distance from root,
\begin{equation}
 \Dout = 
\sum_{i=1}^\infty \Bin(1+D^{(i)},p(\xi_i)) \label{Douteq3}
\end{equation}
which becomes (\ref{Douteq}).

We now turn to the issue of using (\ref{Douteq3}) to get information
about the distribution of $\Dout$.
Because a directed edge contributes equally to total in-degree and to total out-degree, we know a priori that $E\Dout$ must equal $E\Din$, but let us
first check that we can indeed use (\ref{Douteq}) to show
$E\Dout = \beta/(1-\beta)$.
Because
(\ref{bin-mean})
$E \Bin(n,p) = np$
we see
\[ E \Bin(1+D_i,p(\xi_i)) = (1+E\Dout) \ E p(\xi_i).\]
So (\ref{Douteq3}) gives
\[
 E\Dout = (1+E\Dout) \ \cdot EZ^{(1)} 
= (1+E\Dout) \beta \]
giving $E\Dout = \beta/(1-\beta)$.

{\em Variance.}
The calculation of the variance $\var \Dout$
provides a textbook illustration of the utility of the general
{\em conditional variance formula}
\[
\var X = E \var(X|Y) + \var E(X|Y) . \]
We give the details in the low density case;
the high density case is similar.
In the defining equation (\ref{Douteq})
write $D$ for $\Dout$ and write
$\bD$ and $\bxi$ for the random sequences
$(D_i)$ and $(\xi_i)$.
Because
(\ref{var-Bin})
$\var \Bin(n,p) = np(1-p)$
we have
\[ \var(D|\bD, \bxi)
= \sum_i (1+D_i) 
\alpha \lambda
e^{-\lambda \xi_i}
(1 - \alpha \lambda e^{-\lambda \xi_i} ) . \]
Recursive self-similarity, as used above,
implies independence of the i.i.d. sequence $(D_i, \ i \geq 1)$
and the Poisson process $(\xi_i, \ i \geq 1)$.
So
\begin{eqnarray}
E \var(D|\bD, \bxi)&=&
(1+ED) (EZ^{(1)} - EZ^{(2)})\nonumber\\
&=& \frac{\alpha
(1 - \frac{\alpha \lambda}{2})}
{1 - \alpha} 
\mbox{ using (\ref{EDeq}) and (\ref{EZZ})}.
\label{vD1}
\end{eqnarray}
Next consider the conditional expectation
\[
E(D|\bD,\bxi)=
\sum_i (1+D_i)
\alpha \lambda
e^{-\lambda \xi_i}
= W, \quad \mbox{say} . 
\]
We will calculate $\var W$ by using the
conditional variance formula.
Because
\[ \var(W|\bxi)
= \sum_i (\var D) \cdot 
\alpha^2 \lambda^2
e^{-2 \lambda \xi_i} \]
we have
\begin{equation}
E \var(W|\bxi)
= (\var D) \cdot 
EZ^{(2)} =
 (\var D) \cdot 
\alpha^2 \lambda/2
 . \label{vD2}
\end{equation}
And since
$ E(W|\bxi) = (1+ED)
Z^{(1)}
= \sfrac{1}{1-\alpha}Z^{(1)} $
we have
\begin{equation}
\var E(W|\bxi) =
\frac{1}{(1-\alpha)^2}
\var Z^{(1)}
= \frac{\alpha^2 \lambda
}{2(1-\alpha)^2}
. \label{vD3}
\end{equation}
Using the conditional variance formula twice
\begin{eqnarray*}
\var D &=&
E \var(D|\bD, \bxi) 
+ \var W \\
&=&E \var(D|\bD, \bxi) + E \var(W|\bxi) + \var E(W|\bxi)\\
&=& \frac{\alpha
(1 - \frac{\alpha \lambda}{2})}
{1 - \alpha} +
 (\var D) \cdot 
\alpha^2 \lambda/2
+
 \frac{\alpha^2 \lambda
}{2(1-\alpha)^2} .
\end{eqnarray*}
Solving gives the equation (\ref{varDouteq}) for $\var \Dout$.

{\em Special cases.}
{\bf (a).}
Fix $\alpha$.
Because $p(x) \leq \alpha \lambda$,
in the $\lambda \to 0$ limit we can apply the Poisson limit
of Binomials result to
the defining equation (\ref{Douteq}) to obtain
(cf. section \ref{sec-extreme})
\begin{equation}
\mbox{(in $\lambda \to 0$ limit)}
\quad \quad \quad 
\Dout \ed \Poi(\alpha) .
\end{equation}

\noindent
{\bf (b).}
Fix $\eta$.
In the $\lambda \to \infty$ limit
we can use the
limit process of section \ref{sec-extreme}
to show that $1+\Dout$ has the distribution of the 
total population size in a 
Galton-Watson branching process\footnote{A population process starting with one individual in generation $0$, individuals having i.i.d. random numbers of offspring in successive generations}
with $\Poi(\eta)$ offspring distribution.
This is (see e.g. \cite{me79}) the {\em Borel-Tanner}$(\eta)$
distribution
\begin{equation}
\mbox{(in $\lambda \to \infty$ limit)}
\quad \quad \quad 
P(1+\Dout = d) =
\frac{(\eta d)^{d-1} e^{-\eta d}}{d!}, \quad d \geq 1 .
\end{equation}

\noindent
{\bf (c).}
In the case $\alpha \lambda = 1$ 
it turns out (an argument is sketched in section \ref{sec-outcrit})
\begin{equation}
1+\Dout \ed \Geo(1-\alpha) .
\label{eq-outcrit}
\end{equation}

{\em Independence of in-degree and out-degree}.
This independence, noted at (\ref{DDind}),
follows from the fact that in the
forwards space-time PWIT process
the out-degree of the root is determined at the arrival time 
of the root vertex;
the subsequent evolution of the process
of in-edges is clearly independent of the state of the graph
immediately after arrival.

\subsection{Densities of induced subgraphs}
\label{sec-induced-2}
Here we give details of the definition and interpretation of
``density of induced subgraphs" mentioned in section \ref{sec-induced-1},
and list explicit formulas.

Let $G$ and $\GG$ be finite directed acyclic graphs;
think of $G$ as small and $\GG$ as large.
Define
``density of $G$ as an induced subgraph of $\GG$"
by
\[
\dens(G|\GG) =
\frac{
\#\{V \subset \GG: \ V \mbox{ isomorphic to } G\}
}{
\# \{\mbox{ vertices of } \GG\} }
\]
where $\#$ denotes cardinality (``number of") and $V$ denotes a
vertex-subset of $\GG$ with its induced subgraph.
See Figure 7, where there are 3 such vertex-subsets
$\{a,b,e\}, \ \{b,c,e\}, \ \{c,d,e\}$
and so where
$\dens(G|\GG) =
3/5$.

\setlength{\unitlength}{0.34in}
\begin{picture}(12,2.4)(-0.3,0.2)
\put(0,2){\circle*{0.2}}
\put(2,2){\circle*{0.2}}
\put(4,2){\circle*{0.2}}
\put(6,2){\circle*{0.2}}
\put(8,2){\circle*{0.2}}
\put(9,2){\circle*{0.2}}
\put(11,2){\circle*{0.2}}
\put(1,1){\circle*{0.2}}
\put(3,1){\circle*{0.2}}
\put(7,1){\circle*{0.2}}
\put(10,1){\circle*{0.2}}
\put(0.2,1.8){\vector(1,-1){0.6}}
\put(2.2,1.8){\vector(1,-1){0.6}}
\put(6.2,1.8){\vector(1,-1){0.6}}
\put(9.2,1.8){\vector(1,-1){0.6}}
\put(1.8,1.8){\vector(-1,-1){0.6}}
\put(3.8,1.8){\vector(-1,-1){0.6}}
\put(7.8,1.8){\vector(-1,-1){0.6}}
\put(10.8,1.8){\vector(-1,-1){0.6}}
\put(0.2,2){\vector(1,0){1.6}}
\put(3.8,1.9){\vector(-3,-1){2.6}}
\put(-0.09,2.14){$a$}
\put(1.91,2.14){$c$}
\put(3.91,2.14){$e$}
\put(0.92,0.58){$b$}
\put(2.92,0.58){$d$}
\put(2,0.4){$\GG$}
\put(6.4,2.1){$G_1^*$}
\put(9.4,2.1){$G_2^*$}
\put(7.2,0.87){root}
\put(11.15,1.9){root}
\end{picture}

{\bf Figure 7.}
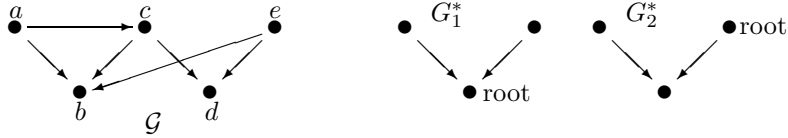
{\small Two rootings $G^*_1, G^*_2$ of a graph $G$.}

\vspace{0.12in}
\noindent

We want to study $n \to \infty$ limits of 
$\dens(G|\GG_n) $
in our MFSC random graph model $(\GG_n)$, for fixed $G$.
To use our methodology we must first rephrase
the definition of
$\dens(G|\GG) $
in terms of the rooted graph $\GG^*$ obtained by giving
$\GG$ a uniform random root.
For such $\GG^*$, and for an arbitrarily-rooted directed graph $G^*$,
define a random variable
\begin{quote}
$\bX(G^*|\GG^*) = $
number of vertex-subsets
$V$ of $\GG^*$ including the root such that
$\GG^*$ restricted to
$V$ is isomorphic to $G^*$ via a root-preserving isomorphism.
\end{quote}
The randomness arises only from choice of root of $\GG^*$;
notation $\bX$ has no special significance except to
distinguish this from simpler random variables.

It is an easy fact that
\begin{equation}
\dens(G|\GG) = 
\frac{E \bX(G^*|\GG^*)}{\iota (G^*)} 
\label{dens-identity}
\end{equation}
where $G^*$ is $G$ with an arbitrary choice of root, and
where $\iota(G^*) \geq 1$ is the number of different
root-choices which would give a rooted graph isomorphic to
this particular choice.
Rather than write a formal proof, let us just
illustrate identity (\ref{dens-identity}) using Figure 7.
For the choice of root giving $G^*_1$, the number of isomorphic
vertex-subsets $V$ of $\GG^*$ equals
$2$ (resp. $1$) if the root of $\GG^*$ chances to be $b$ (resp. $d$),
and so
$E \bX(G^*_1|\GG^*) = 3/5$.
For the choice of root giving $G^*_2$, the number of isomorphic
vertex-subsets $V$ of $\GG^*$ equals
$1$ (resp. $2,3$) if the root of $\GG^*$ chances to be $a$ (resp. $c,e$),
and so
$E \bX(G^*_2|\GG^*) = 6/5$.
Since $\iota(G^*_1) = 1$ while $\iota(G^*_2) = 2$, 
we have checked identity (\ref{dens-identity}) in this example.

Note that in Figure 7, the induced subgraph on
$\{a,b,c\}$ is not isomorphic to $G$ because of the extra edge
$a \to c$.
Obviously we can make parallel definitions allowing
extra edges
(pedantically: replace ``isomorphism" by ``vertex-bijection and
edge-surjection")
and we write
$\bdens(G|\GG)$
and $\bar{\bX}(G^*|\GG^*)$
in this setting.
For instance, in Figure 7 we have
$\bdens(G|\GG) = 4/5$
and $E\bar{\bX}(G_1^*|\GG^*) = 4/5$.

The point of all this is that the definition of
$\bX(G^*|\GG^*)$
makes sense when $\GG^*$ is a rooted {\em infinite} graph.
The key methodology in our analysis of the MFSC model
$(\GG_n)$ is that the randomly-rooted $\GG^*_n$ converge
locally to a limit random infinite rooted graph $\GG^*_\infty$,
implying
via (\ref{dens-identity}) that
\begin{equation}
\dens(G|\GG_n) \to 
\frac{E \bX(G^*|\GG^*_\infty)}{\iota(G^*)}
:= \dens_\infty (G), \mbox{ say}. 
\label{dens-n}
\end{equation}
In parallel.
\begin{equation}
\bdens(G|\GG_n) \to 
\frac{E \bar{\bX}(G^*|\GG^*_\infty)}{\iota(G^*)}
:= \  \bdens_\infty (G), \mbox{ say}. 
\label{dens-nb}
\end{equation}

In sections \ref{sec-dens} and \ref{sec-K22} we calculate $\dens(G)$
for several cases of $G$; let us record the formulas below.
Note that limit densities may be infinite, in which cases we will
point out the conditions on parameters needed for finiteness.

\noindent
{\bf (a).}
For a directed path $\pi_r$ with $r \geq 1$ edges,
\begin{equation}
\bdens_\infty(\pi_r) = \partial^r .
\label{chi-path}
\end{equation}

\noindent
{\bf (b).}
For the complete directed acyclic graph $K_r$
on $r \geq 2$ vertices,
(that is, vertices $\{1,2,\ldots,r\}$ and edges $i \to j$
for $1 \leq i < j \leq r$),
\begin{equation}
\dens_\infty(K_r) = 
\prod_{u=1}^{r-1} \frac{\beta_u}{1 - \beta_u} 
\label{chi-K}
\end{equation}
where
$\beta_1 = \beta$ and for general $u \geq 1$
\begin{equation}
\beta_u := 
\left\{ \begin{array}{cc}
u^{-1} \alpha^u \lambda^{u-1} & \mathrm{[low]} \\
\eta + \sfrac{1}{u \lambda} & \mathrm{[high]}.
\end{array} \right.
\label{EZ}
\end{equation}
In particular, for the case of {\em triangles} $K_3$
we have explicitly
\begin{equation}
 \dens_\infty(K_3) = \left\{ \begin{array}{cc} \frac{\alpha^3 \lambda}{(1-\alpha)(2 - \alpha^2 \lambda)} 
& \low \\
 \frac{(\eta + \frac{1}{\lambda})(\eta + \frac{1}{2\lambda})}
{(1-\eta - \frac{1}{\lambda})(1 - \eta - \frac{1}{2\lambda})}
& \high .
\end{array} \right.
\label{tri-mean}
\end{equation}
As already mentioned in section \ref{sec-induced-1},
the formula above is the key ingredient in the formula for $\ccluster$.
Recall its verbal description
\begin{quote}
The proportion of directed $2$-paths
$v_1 \to v_2 \to v_3$ 
for which $v_1 \to v_3$ is also an edge.
\end{quote}
This becomes
\begin{equation}
\ccluster = 
\frac{\dens_\infty(K_3)}{\bdens_\infty(\pi_2)}
\label{tri-form}
\end{equation}
and then (\ref{chi-path},\ref{tri-form}) immediately give the formula
(\ref{clust-form}).

\noindent
{\bf (c).}
For a directed path $\pi_r$ with $r \geq 1$ edges,
\begin{equation}
\dens_\infty(\pi_r) = \delta 
\left( \frac{\beta_1 - \beta_2}{(1-\beta_1)(1-\beta_2)}\right)^{r-1} .
\label{path-not}
\end{equation}

\noindent
{\bf (d).}
For the complete bipartite directed graph $K_{2,2}$,
for $\beta_2 < \frac{1}{2}$
(which always holds in the low density case)
\begin{equation}
 \bdens_\infty(K_{2,2}) =
\frac{\partial \beta_2(\beta_2 + \sfrac{1}{2}\partial\beta)}
{(1-2\beta_2)(1-\beta_2)} .
\label{K22dens}
\end{equation}

\noindent
{\bf (e).}
In principle one can calculate 
$\dens_\infty(G^*)$ for any $G^*$, but in practice 
it is not clear to what extent useful explicit formulas can be found -- see
section \ref{sec-dens} for further discussion, and for the observation that 
certain graphs $G^*$  
have $\dens_\infty(G^*) = 0$, ``asymptotically negligible density".

\subsection{Densities of induced subgraphs: easy explicit formulas}
\label{sec-dens}
Recall the basic result (\ref{dens-n}) on subgraph density:
\[
\dens(G|\GG_n) \to 
\frac{E \bX(G^*|\GG^*_\infty)}{\iota(G^*)}
:= \dens_\infty (G) \] 
where $G^*$ is an arbitrary rooting of $G$.
In calculating the right side, to simplify notation we write
\[ \chi(G^*) = E \bX(G^*|\GG^*_\infty) \]
and similarly for $\bchi(\cdot)$ and $\bar{\bX}(\cdot)$ and $\bdens(G)$.

First consider $\pi_r$, the directed path with $r$ edges,
rooted at the last-arriving vertex,
which we will call the {\em head}.
Clearly
$\chi(\pi_1) = E\Dout = \partial$.
Let us write out the (rather obvious) inductive argument
for calculating $\bchi(\pi_r)$.
Whether or not the root vertex of $\GG_\infty^*$ is the head
of a $r$-path is determined at its arrival time.
Consider a \nn\ $v_i$ of the root, at distance $\xi_i$.
The expected number of ${r-1}$-paths headed by $v_i$
equals $\bchi(\pi_{r-1})$.
So the expected number of $r$-paths of the form
$\root \to v_i \to \ldots$
equals
$\bchi(\pi_{r-1}) \times P(
(\root,v_i) \mbox{ is edge of } \GG_\infty^*)$.
Summing over $i$ gives
\[ \bchi(\pi_r) = \bchi(\pi_{r-1}) \times E\Dout
= \bchi(\pi_{r-1}) \times \partial \]
and so $\bchi(\pi_r) = \partial^r$ by induction.
This is formula (\ref{chi-path}).

The result for the complete directed graph $K_r$ on $r$ vertices
is similar.
For $r=2$ we have
$\chi(K_2) = \chi(\pi_1) = \partial$
and so to establish formula (\ref{chi-K})
by induction it is enough to show
\begin{equation}
\chi(K_{r+1}) = \chi(K_r) \times \frac{\beta_r}{1 - \beta_r} .
\label{Ktoshow}
\end{equation}
In the forwards space-time PWIT process, consider a vertex-set $S_r$
isomorphic to $K_r$, headed by its latest-arriving vertex $v_*$.
At time $t$ after the arrival of $v_*$, let
$N_t = 1 +$ the number of $K_{r+1}$-subgraphs of the forwards space-time PWIT process
which are of the form $\{v\} \cup S_r$ for some $v$;
regard the ``$+1$" as counting $S_r$ itself.
Then $N_t$ is a Yule process of rate
\begin{equation}
 E \sum_i p^r(\xi_i)
= \beta_r \label{bbb}
\end{equation}
because for each vertex $v$ counted in $N_t$, 
a new vertex $v^\prime$ arriving with \nn\ $v$ at 
distance $x$ has chance
$  \left[ p(x) \right]^r$
to create the $r$ edges needed to make $\{v^\prime\} \cup S_r$
be a $K_{r+1}$ subgraph.
Moreover these are the only ways in which a new $K_{r+1}$
of the form $\{v^\prime\} \cup S_r$ can be formed.
By the Yule formula (\ref{Yule-PWIT})
$ N_t \ed \Geo(\exp(-t \beta_r)) $.
Now regard $K_{r+1}$ as rooted by its second-latest arriving vertex.
In $\GG_\infty^*$ the root has age $T \ed \Exp(1)$.
At its arrival time the root headed some random number of
$K_r$'s, with mean $\chi(K_r)$, so by considering the mean
number of $K_{r+1}$'s at the present time
\begin{eqnarray*}
\chi(K_{r+1}) &=& \chi(K_r) \times (EN_T -1) \\
&=& \chi(K_r) \times (E \exp(T \ \beta_r) \ -1)\\
&=& \chi(K_r) \times \left( \frac{1}{1-\beta_r} \ -1 \right)
\end{eqnarray*}
giving (\ref{Ktoshow}).

{\em Other subgraphs.}
The derivations of formulas (\ref{path-not},\ref{K22dens}) 
dealing with paths and $K_{2,2}$ are relegated to
sections \ref{sec-Y2} and 
\ref{sec-K22}.

For the graph
$\mathrm{out-star}_r$
consisting of $r$ out-edges at a root, it is clear that
\[ \bchi(
\mathrm{out-star}_r
) = E { \Dout \choose r} \]
and similarly
\[ \bchi(
\mathrm{in-star}_r
) = E { \Din \choose r} , \]
and these can in principle be evaluated using (\ref{Douteq},\ref{Dineq}).

{\em Some subgraphs have density zero.}

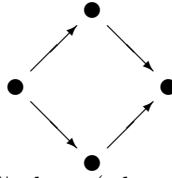
\setlength{\unitlength}{0.4in}
\begin{picture}(2,2.2)(-5,-1)
\put(0.2,0.2){\vector(1,1){0.6}}
\put(0.2,-0.2){\vector(1,-1){0.6}}
\put(1.2,0.8){\vector(1,-1){0.6}}
\put(1.2,-0.8){\vector(1,1){0.6}}
\put(0,0){\circle*{0.2}}
\put(1,1){\circle*{0.2}}
\put(1,-1){\circle*{0.2}}
\put(2,0){\circle*{0.2}}
\end{picture}

It is easy to see that the graph
$G^*$
above (where no ``vertical" edge is present)
has $\chi(G^*) = 0$.

\subsection{Reparametrization}
\label{sec-reparameter}
Writing $\alpha, \lambda$ in terms of $\partial,\kappa = \ccluster$
by solving (\ref{EDeq},\ref{clust-form})
gives the formulas
\begin{equation}
\left. \begin{array}{ccc}
\alpha &=&\frac{\partial}{\partial+1}\\
\lambda&=&\frac{2(1+\frac{1}{\partial})^2}{1+\frac{1}{\partial\kappa}}
\end{array}
\right\}
0<\kappa \leq \frac{1}{\partial+2}
\end{equation}
\begin{equation}
\left. \begin{array}{ccc}
\eta&=&\frac{\partial((\partial+2)\kappa -1)}{(\partial+1)(1+\partial\kappa)}\\
\lambda&=&\frac{(\partial+1)(1+\partial\kappa)}{2 \partial(1-\kappa)}
\end{array}
\right\}
\frac{1}{\partial+2} < \kappa <1 .
\end{equation}

\subsection{Edge-lengths}
The previous calculations have not 
made very extensive use of the time-dynamics of the
forwards space-time PWIT process,
and in particular have not used the fact that edge-lengths
grow exponentially at rate $1$.
To derive the formula (\ref{len-dens}) for edge length density
we do need to exploit such time-dynamics.
We consider only the low-density case; the high density case is
more complicated because the distribution in (ii) below
is no longer exponential.

Consider the lengths of the in-edges at a particular 
vertex $v_0$.
Following a tradition in mathematical probability,
we visualize an in-edge of length $\ell$ as a ``particle"
at position $\ell$ on a line; we also put a particle at position
$0$ to represent the vertex $v_0$ itself.
If we start time $\tau$ with $\tau = 0$ at the arrival time
of $v_0$, 
then the evolution of the ``particle process" can be specified
as follows.

\noindent
(i) There is a particle at position $0$ at all times 
$\tau \geq 0$.

\noindent
(ii) For each particle (at position $x$ at time $\tau$, say),
at stochastic rate $\alpha$ per unit time a new particle
appears at position $x + \Exp(\lambda)$.

\noindent
(iii) particle positions increase deterministically at
exponential(1) rate: a particle at $x$ at time $\tau$
will be at $x e^{\tau_0 - \tau}$ at time $\tau_0 > \tau$.

\noindent
Rule (ii) derives from (\ref{newvertex}):
for an existing edge $(v^\prime,v_0)$,
a new vertex arriving at distance $r$ from \nn\ $v^\prime$
creates an edge to $v_0$ with probability $p(r)$,
so the rate at which each existing edge is copied equals
$\int_0^\infty p(x) \ dx = \alpha$;
moreover conditional on copying, the distance $r$ has
$\Exp(\lambda)$ distribution, and so the length of the new edge
equals the length of the old edge $+ \Exp(\lambda)$.

To analyze this particle process of edge lengths,
define
\[ G(\tau,x) =
E ( \mbox{number of edges of length $>x$ at time $\tau$})
\]
so that
\[ g(\tau,x) = - \frac{d}{dx} G(\tau,x) =
\mbox{mean edge-length density at time $\tau$}.\]
We shall study
\[ f(x) \ dx = 
E (\mbox{number of in-edges at a typical vertex with length
$\in [x,x+dx]$}) .\]
Because the age of a typical vertex has $\Exp(1)$ distribution,
$f(x)$ can be written as
\[ f(x) = \int_0^\infty g(\tau,x) e^{-\tau} \ d \tau . \]
The verbal description of the particle process leads
to the equation
\[
\frac{d}{d\tau} G(\tau,x) =
x g(\tau,x)
+ \alpha \int_0^x g(\tau,y)e^{-\lambda (x-y)} \ dy
+ \alpha e^{-\lambda x}
. \]
Here the first term on the right expresses the deterministic exponential growth,
the second term expresses birth of particles to parents not at $0$ (copying of existing edges)
and the third expresses
births to the $0$-particle (new edge to $v_0$).
Multiply the terms of the equation by $e^{-\tau}$
and integrate out $\tau$; noting
\[ \int_0^\infty
\frac{d}{d\tau} G(\tau,x) e^{-\tau} \ d\tau
= \int_0^\infty G(\tau,x) e^{-\tau} \ d\tau
= F(x), \mbox{say},\]
we obtain
\begin{equation}
 F(x) = xf(x) +
\alpha \int_0^x f(y)e^{-\lambda (x-y)} \ dy
+  \alpha e^{-\lambda x} . \label{Fxf}
\end{equation}
Differentiate:
$ - f = (xf)^\prime + 
\alpha f
- \lambda \alpha \int_0^x f(y)e^{-\lambda (x-y)} \ dy
- \lambda  \alpha e^{-\lambda x} $.
\\
Rewrite with the integral term on the left,
and then substitute the integral term by the expression
implied in (\ref{Fxf}):
\[ \lambda 
(-xf - \alpha e^{-\lambda x} + F)
= f + (xf)^\prime + \alpha f
- \lambda  \alpha e^{-\lambda x} . \]
Differentiate:
$ \lambda (
-(xf)^\prime 
+ \lambda  \alpha e^{-\lambda x} 
-f) = 
f^\prime + (xf)^{\prime \prime}
+ \alpha f^\prime 
+ \lambda^2  \alpha e^{-\lambda x} 
$.
\\
Tidy:
$ (xf)^{\prime \prime}
+ \lambda (xf)^\prime
+ (1+\alpha)f^\prime
+ \lambda f = 0 $.
\\
Look for a series solution
$ f(x) = \sum_{n=0}^\infty a_nx^n $.
Equating coefficients of $x^n$:
\[ (n+2)(n+1)a_{n+1}
+ \lambda (n+1)a_n
+(1+\alpha)(n+1)a_{n+1}
+ \lambda a_n = 0 .\]
That is,
\[ \frac{a_{n+1}}{a_n} = 
\frac{-\lambda (n+2)}
{(n+1)(n+3+\alpha)} \]
and so
\[ a_n = 
\frac{(-\lambda)^n (n+1) \Gamma(3 + \alpha)}{\Gamma(n+3+\alpha)} a_0 . \]
One can directly check that $f(0+) = 1$, identifying $a_0 = 1$.
Because the mean in-degree is $\alpha/(1-\alpha)$,
the probability density function of a typical edge-length must be
$\frac{1-\alpha}{\alpha} f(x)$,
establishing (\ref{len-dens}).

\section{Further calculations}
\label{sec-furtherc}
\subsection{Yule arguments for subgraph density}
\label{sec-Y2}
The next Lemma abstracts the Yule process arguments used in section \ref{sec-dens}.
Recall the reformulation there of limit subgraph density
in terms of $\chi(G)$ and $\bchi(G)$.

\setlength{\unitlength}{0.35in}
\begin{picture}(16,5.8)(-1,-0.9)
\put(0,3){\circle*{0.2}}
\put(0,1){\circle*{0.2}}
\put(0,2.8){\vector(0,-1){1.6}}
\put(2,0){\circle*{0.2}}
\put(0.1,2.86){\vector(2,-3){1.8}}
\put(4,1){\circle*{0.2}}
\put(3.8,0.93){\vector(-2,-1){1.6}}
\put(0.2,2.9){\vector(2,-1){3.6}}
\put(-0.9,2.91){root}
\put(2,3){$G_0$}
\put(6,3){\circle*{0.2}}
\put(6,1){\circle*{0.2}}
\put(6,2.8){\vector(0,-1){1.6}}
\put(8,0){\circle*{0.2}}
\put(6.1,2.86){\vector(2,-3){1.8}}
\put(10,1){\circle*{0.2}}
\put(9.8,0.93){\vector(-2,-1){1.6}}
\put(6.2,2.9){\vector(2,-1){3.6}}
\put(5.1,2.91){root}
\put(9,3){$G_1$}
\put(8,4){\circle*{0.2}}
\put(8.2,4.1){$w$}
\put(7.8,3.93){\vector(-2,-1){1.6}}
\put(8,3.8){\vector(0,-1){3.6}}
\end{picture}
{\bf Figure 8.}
{\small 
Illustration of Lemma \ref{Lnew}.}

\begin{Lemma}
\label{Lnew}
Let $G_0$ be a rooted 
directed acyclic graph
such that each vertex is a descendant of the root.
Let $q$ be the out-degree of the root.
Let $G_1$ be a 
directed acyclic graph
obtained from $G_0$ by adding an extra vertex $w$ and
edges $(w,\root)$ and $c$ further edges from $w$ to some children
of the root (so $0 \leq c \leq q$).
Then
\begin{eqnarray}
\bchi(G_1)/\bchi(G_0) &=&
\frac{\beta_{c+1}}{1-\beta_{c+1}} . 
\label{chirec1} \\
\chi(G_1)/\chi(G_0) &=&
\sum_{j=0}^{q-c} (-1)^j 
{q-c \choose j}
\frac{\beta_{c+1+j}}{1-\beta_{c+1+j}} .
\label{chirec2} \\
\end{eqnarray}
In particular, if $q=c$ then
\begin{equation}
\chi(G_1)/\chi(G_0) =
\frac{\beta_{c+1}}{1-\beta_{c+1}} . 
\label{chirec3} 
\end{equation}
\end{Lemma}
As a quick application let us derive formula
(\ref{path-not})
for $\chi(\pi_r)$
for the directed path $\pi_r$ on $r$ edges.
Applying (\ref{chirec2}) with $q=1, c=0$,
\[ \chi(\pi_{r+1})/\chi(\pi_r) = 
\frac{\beta_1}{1-\beta_1}
- \frac{\beta_2}{1-\beta_2}
= \frac{\beta_1 - \beta_2}{(1-\beta_1)(1-\beta_2)} .\]
Because $\chi(\pi_1) = \partial$
and $\iota(\pi_r) = 1$
we obtain
formula (\ref{path-not}).

{\em Proof of Lemma \ref{Lnew}.}
We will do the harder case (\ref{chirec2}).
Consider a copy of $G_0$ (i.e. an isomorphic subgraph) at the root of the PWIT.
In the space-time PWIT process, let
$M(t)$ be the number of copies of $G_1$ which contain the
given copy of $G_0$,
at time $t$ after the arrival of the root.
Since the age $T$ of the root has $\Exp(1)$ distribution,
\[ \chi(G_1)/\chi(G_0) = EM(T) . \]
Write $S$ for the set of children of the root in $G_0$
and write $A$ for a subset of $S$.

Consider the process of arriving vertices $v$ which form an edge
to the root.
Such a $v$ has a \nn\ $v^\prime$,
where either $v^\prime = \root$ or $(v^\prime,\root)$ is already an edge.
Writing $A(v^\prime) \subseteq S$
for the set of children of the root to which $v^\prime$ creates an edge,
then $A(v) \subseteq A(v^\prime)$.
We can now write
\[ M(t) = \# \{v: \ (v,\root)
\mbox{ is an edge, }
A(v) = A_1 \} \]
where $A_1$ is the set of children of the root of $G_1$ to which 
$w$ has an edge.
In representing $M(t)$ as above, we are using the hypothesis
``each vertex is a descendant of the root"
to ensure that, in a subgraph of the space-time PWIT isomorphic to $G_1$,
the last-arriving vertex must be $w$.

Consider a sequence
$\root = v_{(0)},v_{(1)},\ldots,v_{(i)}$
of arriving vertices such that each vertex $v_{(k)}$
arrives at distance $x_k$ from its \nn\ $v_{(k-1)}$.
The chance that each $v_{(k)}$ makes an edge to the root
and to each child in $A_1$ equals
$\prod_{k=1}^i p^{c+1}(x_k)$.
The chance that furthermore no other child in $S$ acquires an edge to $v_{(i)}$ equals
$(1 - \prod_{k=1}^i p(x_k))^{q-c}$.
By considering the times
$0<t_1<t_2 < \ldots <t_i < t$
of arrivals of $v_{(i)}$,
\[ EM(t) = 
\sum_{i=1}^\infty
\int_{0<t_1< \ldots < }
\int_{ t_i<t}
dt_1 \ldots dt_i
\int_0^\infty
\ldots
\int_0^\infty
dx_1 \ldots dx_i
\ \bp^{c+1}(\bx) (1-\bp(\bx))^{q-c}
\]
where $\bp(\bx) = \prod_{k=1}^i p(x_i)$.
Because 
$\bp^{c+1}(\bx) (1-\bp(\bx))^{q-c}
= \sum_{j=0}^{q-c} (-1)^j {q-c \choose j} \bp^{c+1+j}(\bx)
$ and
$\int_0^\infty p^{c+1+j}(x_k) \ dx_k = \beta_{c+1+j}$,
\begin{eqnarray*}
 EM(t) &=& 
\sum_{i=1}^\infty
\frac{t^i}{i!} 
\sum_{j=0}^{q-c} (-1)^j {q-c \choose j} \beta^i_{c+1+j} \\
&=&
\sum_{j=0}^{q-c} (-1)^j {q-c \choose j} 
(\exp(\beta_{c+1+j}t) -1) .
\end{eqnarray*}
Calculating
$EM(T) = \int_0^\infty e^{-t} EM(t) \ dt$
establishes (\ref{chirec2}).

\subsection{Subgraph density of $K_{2,2}$}
\label{sec-K22}
We have not pursued general methods for induced subgraph density beyond
Lemma \ref{Lnew}, but the argument that follows for the
particular case of $K_{2,2}$, based on splitting into two cases,
could clearly be applied somewhat more widely.

We first quote
\begin{Lemma}
\begin{eqnarray}
E {\mathrm{Geo}(p) \choose 2} &=& p^{-2} - p^{-1} \label{Nc21}\\
E {\mathrm{Geo}(p) - 1 \choose 2} &=& p^{-2} -2p^{-1} + 1 . \label{Nc22}
\end{eqnarray}
\end{Lemma}
Next consider the graph $G^*$ on the left of Figure 9.

\setlength{\unitlength}{0.6in}
\begin{picture}(7,2.5)(0,-1.1)
\put(4.1,0.5){\vector(4,1){1.8}}
\put(4.1,-0.5){\vector(4,-1){1.8}}
\put(4.1,-0.5){\vector(4,3){1.8}}
\put(4.1,0.5){\vector(4,-3){1.8}}
\put(5.6,0.1){\vector(1,2){0.3}}
\put(5.6,-0.1){\vector(1,-2){0.3}}
\put(4,0.5){\circle*{0.1}}
\put(4,-0.5){\circle*{0.1}}
\put(6,1){\circle*{0.1}}
\put(5.5,0){\circle*{0.1}}
\put(6,-1){\circle*{0.1}}
\put(5.64,0){root}
\put(6.4,0.4){$\widetilde{G}$}
\put(6.1,1){$v_{-1}$}
\put(6.1,-1){$v_{-2}$}
\put(0.1,0){\vector(1,0){2.8}}
\put(0.1,-0.05){\vector(2,-1){1.8}}
\put(1.1,0.8){\vector(1,-2){0.8}}
\put(1.1,0.95){\vector(2,-1){1.8}}
\put(2.1,-0.9){\vector(1,1){0.8}}
\put(0,0){\circle*{0.1}}
\put(3,0){\circle*{0.1}}
\put(2,-1){\circle*{0.1}}
\put(1,1){\circle*{0.1}}
\put(1.62,-1.2){root}
\put(3.1,0){$v_{-1}$}
\put(-0.6,0){$G^*$}
\end{picture}

\vspace{0.12in}
{\bf Figure 9.}
{\small 
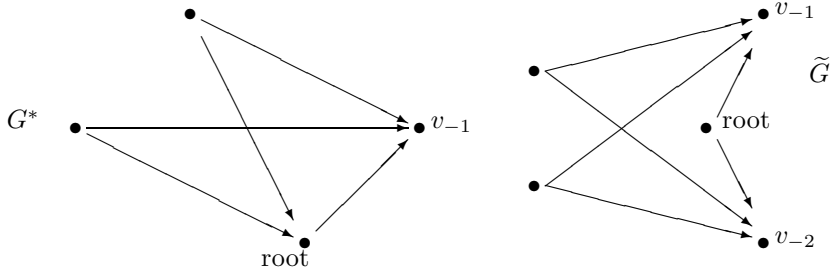
Graphs related to $K_{2,2}$.
}

\vspace{0.1in}
\noindent
We will show
\begin{equation}
\bchi(G^*) = \frac{2\beta_2^2 \partial}{(1-2\beta_2)(1-\beta_2)}
\label{G32}
\end{equation}
where
$\beta_2 = EZ^{(2)}$.

We start by repeating the argument in the $r=2$ case of (\ref{Ktoshow}).
In the 
forwards space-time PWIT process,
consider the newly-arrived root and an edge $(\root,v_{-1})$.
At time $t$ after the arrival of the root, let 
$N_t = 1 +$ the number of 
vertices $v$ such that
$(v,\root)$ and $(v,v_{-1})$ are both edges of the graph process;
regard the ``$+1$" as counting the root itself.
Then $N_t$ is a Yule process of rate
$\beta_2 = \int_0^\infty p^2(x) \ dx$.
Thus at time $t$ there are 
${N_t - 1 \choose 2}$
graphs of the desired form containing the edge $(\root,v_{-1})$.
Because the age $T$ of the root has $\Exp(1)$ distribution,
we see
\[ \bchi(G^*) = \partial E
{N_T - 1 \choose 2} \]
where $\partial = E\Dout$ is the expected number of edges of the form
$(\root,v_{-1})$.
Using (\ref{Yule-1}) and 
(\ref{Nc22}),
\begin{eqnarray*}
 \bchi(G^*) &=& \partial E
\left( e^{2\beta_2T} - 2e^{\beta_2T} + 1 \right) \\
&=& \partial  
\left( \sfrac{1}{1-2\beta_2} - \sfrac{2}{1-\beta_2} +1\right)
\end{eqnarray*}
leading to (\ref{G32}).

Next, in $\GG_\infty^*$ consider
\begin{quote}
$ Q := \mbox{ number of unordered pairs $(v_{-1},v_{-2})$ such that}$
 $(\root,v_{-1})$ and $(\root,v_{-2})$ are edges, and
$v_{-1}$ and $v_{-2}$ were in different geometric components at the
arrival time of the root.
\end{quote}
By considering distances $r_1,r_2$ from the root to the 
\nn s of the geometric components  
containing $v_1,v_2$,
\[ EQ = \sfrac{1}{2} \int \int p(r_1)p(r_2) \ dr_1dr_2 \times \partial^2
= \beta^2 \partial^2/2 . \]

Now consider in $\GG_\infty^*$ configurations  
$\widetilde{G}$ as on the right of Figure 9,
where there is no edge between $v_{-1}$ and $v_{-2}$,
and where the root is the first-arriving vertex to have edges to
both $v_{-1}$ and $v_{-2}$; these requirements are equivalent
to saying that at the arrival time of the root,
$v_{-1}$ and $v_{-2}$ were in different geometric components.
Reuse a now-familiar argument.
At time $t$ after the arrival of the root, let 
$N_t = 1 +$ the number of 
vertices $v \neq \root$ such that
$(v,v_{-2})$ and $(v,v_{-1})$ are both edges of the graph process;
regard the ``$+1$" as counting the root itself.
Then $N_t$ is a Yule process of rate
$\beta_2 = \int_0^\infty p^2(x) \ dx$.
Thus at time $t$  
the number of possible unordered pairs $\{v_1,v_2\}$
which give the configuration in the figure,
{\em where we allow one of $\{v_1,v_2\}$ to be the root},
equals 
${N_t \choose 2}$.
Because the age $T$ of the root has $\Exp(1)$ distribution,
we see
\[ \bchi(\widetilde{G}) =
 E {N_T \choose 2} \times EQ  \]
where $\bchi(\widetilde{G})$ is the density of graphs as
on the right of Figure 9, perhaps with extra edges,
but subject to the requirement that 
the root is the first-arriving vertex to have edges to
both $v_{-1}$ and $v_{-2}$.  
Using (\ref{Yule-1}) and 
(\ref{Nc22}),
\[ E {N_T \choose 2} =
E \left( e^{2\beta_2 T} - e^{\beta_2T} \right)
= \sfrac{1}{1-2\beta_2} - \sfrac{1}{1-\beta_2}
= \frac{\beta_2}{(1-2\beta_2)(1-\beta_2)} 
. \]
One can now write
\[ \bdens(K_{2,2}) = \sfrac{1}{2} \bchi(K_{2,2}) = 
\bchi(G^*) + \bchi(\widetilde{G}) \]
because a $4$-vertex graph in $\GG_\infty^*$ containing $K_{2,2}$
is either of the form $G^*$ or is the restriction of a graph
of the form $\widetilde{G}$, in which the extra root is specified
by the requirement stated above
(the factor $1/2$ reflects the fact $\iota(K_{2,2}) = 2$).
Combining the formulas above gives
\[ \sfrac{1}{2} \bchi(K_{2,2}) =
  \frac{\beta_2^2 \partial}{(1-2\beta_2)(1-\beta_2)}
+ \beta^2 \partial^2/2  \times
 \frac{\beta_2}{(1-2\beta_2)(1-\beta_2)} 
  \]
which simplifies to (\ref{K22dens}).

\subsection{Directed percolation}
\label{sec-DP}
Here we record some calculations without detailed explanation.
In the context of the space-time PWIT and the 
evolving random graph process $\GG_\infty^*(s)$,
we can seek to grow a ``core" graph $\CC(s)$ inside $\GG_\infty^*(s)$
via a greedy rule:
\begin{quote}
a newly-arriving vertex is included in $\CC(s)$ if it creates
an edge to some vertex already in $\CC(s)$, in which case all
such edges are included in $\CC(s)$.
\end{quote}
If this construction works, we expect the process $(\CC(s))$ to have a a stationary distribution
$\CC(0)$, say, where 
$\CC(0) \subset \GG_\infty^*$.
Consider
\begin{eqnarray*}
q &=& P(\root \in \CC(0)) \\
Y &=& \mbox{ out-degree of $\root$ in $\CC(0)$, given $\root \in \CC(0)$} .
\end{eqnarray*}
Consider the relation
\[
\widetilde{Y} = \sum_{i=1}^\infty \Ber_i(q) \Bin_i(1+Y_i,p(\xi_i))
  \]
where we write $\Ber(p)$ for a Bernoulli$(p)$ r.v.
(taking value $1$ with probability $p$ and value $0$ otherwise).
Using the recursive structure of the limit random graph process,
we see that $q$ and $Y$ solve the equations
(for unknown $0<q<1$ and an unknown distribution $Y$ on 
$\{1,2,3,\ldots\}$)
\begin{equation}
Y \ed \mbox{dist}(\widetilde{Y}|\widetilde{Y} \geq 1) ; \quad
q = P(\widetilde{Y} \geq 1) .
\label{dirperceq}
\end{equation}
Define
$
p_{\mbox{\small{dir-perc}}}(\alpha,\lambda)
$ to be the solution $q$ if it exists, and to be $0$ otherwise.
The interpretation of this quantity in terms of the
finite random graph process $(\GG_n, n \geq 1)$ is that
\[ n^{-1} E \mathbf{T} \to 
p_{\mbox{\small{dir-perc}}}(\alpha,\lambda)
\]
where $\mathbf{T}$ is the maximal size of a tree in $\GG_n$
directed toward some root.
So in particular, for
$p_{\mbox{\small{perc}}}(\alpha,\lambda)
$ defined at (\ref{pperc}),
\[
p_{\mbox{\small{dir-perc}}}(\alpha,\lambda)
\leq
p_{\mbox{\small{perc}}}(\alpha,\lambda)
. \]
Equation (\ref{dirperceq}) in principle
determines 
$p_{\mbox{\small{dir-perc}}}(\alpha,\lambda)$,
but to get an explicit bound we reuse an underlying idea.
Because $1+ Y_i \geq 2$,
\[
\widetilde{Y} \geq \sum_{i=1}^\infty \Ber_i(q) \Bin_i(2,p(\xi_i))
= Y^*, \mbox{ say.} \]
If the equation
\begin{equation}
q = P(Y^* \geq 1)
\label{qqY}
\end{equation}
has a solution $q>0$ then one can argue
$p_{\mbox{\small{dir-perc}}}(\alpha,\lambda)
\geq q$.
But (\ref{qqY}) is an explicit equation
\[ 1-q = \exp \left( -
\int (2p(x)-p^2(x))q \ dx \right)
= \exp(-(2\beta - \beta_2)q) . \]
If $2\beta - \beta_2 > 1$ there is a solution $q>0$,
establishing (\ref{dirpercbd}).

\subsection{Out-degree in the case $\alpha \lambda = 1$}
\label{sec-outcrit}
The special property of this case is that $p(x) = e^{-\lambda x}$.
On the PWIT consider
\[ Y = \sum_{v \neq \root} \Ber(e^{-\lambda d(v,\root)}) . \]
This satisfies the same recursion (in the special case)
as does $\Dout$.
But there is another way to study $Y$,
which we sketch briefly.
Either the root of the PWIT has no children within a
small distance $\delta$;
or it does have a child, and the distances to the other descendants
of the root and of this child are independent copies of the PWIT
distances.
Because the effect on $Y$ of increasing distances by $\delta$
is to censor each Bernoulli success with probability $\lambda \delta$,
we see that $Y$ is the stationary distribution of the continuous-time
Markov chain on states $\{0,1,2,\ldots\}$
with dynamics
\begin{eqnarray*}
y \to y -1&:& \mbox{ rate } \lambda y \\
y \to y+ \hat{Y} + 1 &:& \mbox{ rate } 1
\end{eqnarray*}
where $\hat{Y}$ is an independent copy of $Y$.
One can now check algebraically that
\[ P(Y = y) = (1- \sfrac{1}{\lambda})(\sfrac{1}{\lambda})^y, \quad y \geq 0 \]
solves the balance equations for this chain.
That is, $1+\Dout$ has $\Geo(1-\sfrac{1}{\lambda}) = \Geo(1-\alpha)$
distribution, as asserted in
(\ref{eq-outcrit}).

{\em Remark.}  Antar Bandyopadhyay (personal communication) has given a purely
analytic verification of (\ref{eq-outcrit}).

\subsection{Triangle density of a function of degree}
\label{sec-Ck}
Here we outline an argument for (\ref{eq-Ck}).
Because $\Din$ has power-law tail and $\Dout$ has geometric tail,
when $D = \Din + \Dout$ is large, say $k$, then 
$\Dout = O(1)$ and $\Din = k - O(1)$.
It is then not hard to argue that the large-$k$ behavior
for $C(k)$ will be the same as for
\[ C^*(k) = \frac{
E(\mbox{number of triangles with in-vertex $v_0\ | \ v_0$ has in-degree $k$})}
{{k \choose 2}}  \]
where the {\em in-vertex} of a triangle is the vertex with two in-edges.

Recall from section \ref{sec-Din} that 
\begin{quote}
$N(t) = $ number of in-edges at a typical vertex $v_0$ at time $t$
after its arrival
\end{quote}
is the Yule process of rate $\beta$.
Write $v_1,v_2,v_3,\ldots$ for the successive arriving vertices
which create edges to $v_0$, and for $i \geq 2$ write
\[ M_i = \mbox{ number of edges from $v_i$ to $\{v_{i-1},v_{i-2},\ldots,v_1\}$} . \]
After $v_k$ arrives there are
$M_2+M_3+ \ldots +M_k$
triangles with in-vertex $v_0$.
If we can show
\[ EM_k \to b \mbox{ as } k \to \infty \]
then we will have
\begin{equation}
 C(k) \approx \frac{kb}{{k \choose 2}} \sim \frac{2b}{k} . \label{CK9}
\end{equation}
Here we are sliding over the fact that $1+\Din$ is the Yule process
evaluated at an independent $\Exp(1)$ time $T$; conditioning 
this to take a value $k$ does not affect the properties used in the
argument below.

Suppose vertices $v_1,\ldots,v_{k-1}$ have arrived and consider
what edges will be created when $v_k$ arrives.
The dynamics (\ref{newvertex}) of the space-time PWIT
say
\begin{quote}
the rate of arrival of new vertices with some one of
$v_0,v_1,\ldots,v_{k-1}$ as \nn\ and at distance
$\in [x,x+dx]$ from that \nn\ equals
$k \ dx$.
The index $I$ of that \nn\ $v_I$
is uniform on $\{0,1,\ldots,k-1\}$.
\end{quote}
Such an arriving vertex creates an edge to $v_0$
with probability $p(x)$.
So conditional on that event (meaning the arriving vertex is $v_k$),
the distance $\hat{\xi}$ from the \nn\ $v_I$ and the index $I=I_k$
of that \nn\ satisfy\\
(i) $\hat{\xi}$ has probability density function 
$\hat{p}(x) = \frac{p(x)}{\int p(u) du}$;\\
(ii)
$I$ is uniform on $\{0,1,\ldots,k-1\}$.\\
Because $v_k$ will copy each of the $M_I$ out-edges from $v_I$
with probability $p(\xi)$ each, and create an edge to $V_I$
with the same probability, 
we obtain the recursion
\[ M_k \ed \Bin(1+M_I,p(\hat{\xi})) \]
where $M_1 = 0$ and where we interpret the right side as $0$
when $I = 0$.
So the limit $\lim_k EM_k = b$ solves
$ b = (1+b) Ep(\hat{\xi}) $
and so
$ b = \frac{Ep(\hat{\xi})}{1 - Ep(\hat{\xi})} $.
Finally,
\[ Ep(\hat{\xi}) = \int p(x) \hat{p}(x) \ dx
= \beta_2/\beta . \]
So $b = \frac{\beta_2}{\beta - \beta_2}$
and (\ref{eq-Ck}) follows 
from (\ref{CK9}).

\section{Comparison with other models}
\label{sec-compare}

Recent complex networks models fall into two categories.
In the {\em small worlds} models popularized by Strogatz and Watts,
vertices are points in $d$-dimensional space, which automatically provides a metric distance between vertices, and the model
uses some rule to
create a random graph with short-range and long-range edges.
In purely graph-theoretical models, such as the 
basic {\em proportional attachment} model popularized by
Albert and Barab{\'a}si\footnote{but really just a minor variation
of Yule's idea: see Lecture 4 of \cite{me-Net}}
the vertices have no ``intrinsic structure" other than that provided
by the graph; we visualize this as saying that each pair of vertices
is metric distance $1$ apart.
In a metric copying model we visualize vertices
as points in some abstract metric space, representing
(in the case of web pages, say) the difference between the
content of the pages, or 
(for people) some notion of ``social distance" based on location, education, 
profession, interests etc of the individuals.
In detail the mean-field model of distance model is used for mathematical tractability rather
than any claimed realism.
But the exponential growth of number of vertices with metric distance
is intermediate between,
and surely in many contexts more plausible than,
the alternatives implicit in the two standard categories of model above.

Within graph-theoretic models, the idea of distance preferences
in attachment has been explored (see \cite{JJ02} and citations therein).
But the general idea of combining proportional attachment with 
metric geometry has scarcely been explored\footnote{\cite{menczer02} gives
a simulation study of an explicitly power-law model, as well as
interesting empirical study of a notion of {\em lexical distance}
between web pages}, 
and the specific use of the mean-field model
is novel.

As a technical note, the mean-field model is a 
zero-parameter\footnote{Zero {\em dimensionless} parameters, to be pedantic}
model of distance.
Our full network model has the two parameters $(\alpha,\lambda)$;
in contrast a typical small-worlds network model has four
parameters
(dimension, number of short-range links, constant and exponent
for probability of long-range edges).

As another technical note, the property
(cf. (\ref{eq-Ck}))
$C(k) \sim c/k$
has been proposed \cite{RB03}
as a criterion for identifying networks which are
``hierarchical" in some sense.
But in our finite-$n$ model (recall section \ref{sec-MFSC})
each vertex has qualitatively
the same behavior, rather than different vertices being 
{\em a priori} assigned different hierarchical roles.
So our model is non-hierarchical, and
we are inclined to regard the criterion as 
ineffective\footnote{One could alternatively regard it as
indicating some subtle emergent hierarchical structure;
cf. \cite{TGJ02}}.

The specific model studied in this paper
is intended as a ``general purpose" model rather than being
tuned to some particular subclass of real-world networks.
Having as one ingredient the now-familiar proportional attachment
feature, one could look at the many existing variant models
in the literature and explore them within our platform.
In other words,
there are many ways to add a third parameter 
intended to express some presumed real-world feature
or some theoretical desideratum.
For instance
\begin{itemize}
\item One can impose connectivity by requiring that a new vertex {\em always}
links to its nearest neighbor.
\item one can add rules allowing a new vertex to immediately
acquire in-edges, or for edges to randomly appear between existing edges.
Such rules can be designed
(as in e.g. \cite{boll-scale} section 11) to 
produce power law distributions for in-degree.
\end{itemize}

\subsection{Concluding remarks}
In this paper we have focused on
\begin{itemize}
\item describing the model and its conceptual background
(section \ref{sec-MODEL})
\item listing explicit formulas (sections \ref{sec-2par} -- \ref{sec-EL})
and exhibiting the calculations which lead to these formulas
(sections \ref{sec-CALC} and \ref{sec-furtherc}).
\end{itemize}
We are postponing to a later paper consideration of
\begin{itemize}
\item technical issues in the relation between the finite-$n$
model and its infinite limit $\GG_\infty^*$
\item the open problems indicated in sections \ref{sec-OLS} -- \ref{sec-AD},
whose study requires the ``bounding" techniques of theoretical
mathematical probability rather than explicit calculations.
\end{itemize}

\paragraph{Acknowledgement}
I thank an anonymous referee for detailed constructive comments.

\def\cprime{$'$} \def\polhk#1{\setbox0=\hbox{#1}{\ooalign{\hidewidth
  \lower1.5ex\hbox{`}\hidewidth\crcr\unhbox0}}} \def\cprime{$'$}
  \def\cprime{$'$} \def\cprime{$'$}
  \def\polhk#1{\setbox0=\hbox{#1}{\ooalign{\hidewidth
  \lower1.5ex\hbox{`}\hidewidth\crcr\unhbox0}}} \def\cprime{$'$}
  \def\cprime{$'$} \def\polhk#1{\setbox0=\hbox{#1}{\ooalign{\hidewidth
  \lower1.5ex\hbox{`}\hidewidth\crcr\unhbox0}}} \def\cprime{$'$}
  \def\cprime{$'$} \def\cydot{\leavevmode\raise.4ex\hbox{.}} \def\cprime{$'$}
  \def\cprime{$'$} \def\cprime{$'$} \def\cprime{$'$}

\end{document}